\documentclass[aps,
nofootinbib,
superscriptaddress,
showpacs,twocolumn,
pra,longbibliography]{revtex4-1}
\usepackage{amsmath}
\usepackage{easybmat}
\usepackage{amsfonts}  
\usepackage{enumitem}
\usepackage{graphicx}
\usepackage{amsthm}
\usepackage{subfigure}
\usepackage{array}
\usepackage{hyperref}
\usepackage[usenames]{color}
\usepackage{centernot}

\usepackage{rotating}

\usepackage{braket}

\renewcommand{\v}[1]{\mathbf{#1}}
\renewcommand{\vec}[1]{\v{#1}}

\newcommand{\lp}{\left ( }
\newcommand{\rp}{\right ) }

\newcommand{\hc}{\text{H.c.}}

\newcommand{\beq}{\begin{eqnarray*}}
\newcommand{\eeq}{\end{eqnarray*}}

\newcommand{\be}{\begin{eqnarray}}
\newcommand{\ee}{\end{eqnarray}}

\newcommand{\mc}{\mathcal}

\newcommand{\rvdw}{R_{\text{vdW}}}
\newcommand{\rsr}{r_{\text{sr}}}
\newcommand{\lho}{l_{\text{ho}}}
\newcommand{\Hb}{{\hat H}_{\text{b}}} 

\newcommand{\assignment}[1]{}

\def\lsim{\mathrel{\rlap{\lower4pt\hbox{\hskip1pt$\sim$}}
    \raise1pt\hbox{$<$}}}                

\def\gsim{\mathrel{\rlap{\lower4pt\hbox{\hskip1pt$\sim$}}
    \raise1pt\hbox{$>$}}}                

\begin{document}

\title{
Lattice model parameters for ultracold nonreactive molecules:  chaotic scattering and its limitations
}
\author{Michael L. Wall} 
\thanks{Present address: The Johns Hopkins University Applied Physics Laboratory, Laurel, MD 20723, USA}
\affiliation{JILA, NIST and University of Colorado, Boulder, Colorado 80309-0440, USA}
\author{Rick Mukherjee}  
\affiliation{Department of Physics and Astronomy, Rice University, Houston, Texas 77005, USA}
\affiliation{Rice Center for Quantum Materials, Rice University, Houston, Texas 77005, USA}
\author{Shah Saad Alam}  
\affiliation{Department of Physics and Astronomy, Rice University, Houston, Texas 77005, USA}
\affiliation{Rice Center for Quantum Materials, Rice University, Houston, Texas 77005, USA}
\author{Nirav P. Mehta} 
\affiliation{Department of Physics and Astronomy, Trinity University, San Antonio, Texas 78212, USA}
\author{Kaden R.~A. Hazzard} \email{kaden.hazzard@gmail.com}
\affiliation{Department of Physics and Astronomy, Rice University, Houston, Texas 77005, USA}
\affiliation{Rice Center for Quantum Materials, Rice University, Houston, Texas 77005, USA}

\begin{abstract}
We calculate the parameters of the recently-derived many-channel Hubbard model that is predicted to describe ultracold nonreactive molecules in an optical lattice,  going beyond the approximations used in 
Do\c{c}aj \textit{et al.}~[Phys. Rev. Lett. \textbf{116}, 135301 (2016)].
 Although those approximations are expected to capture  the qualitative structure of the model parameters, finer details and quantitative values are less certain.  To set expectations for experiments, whose results depend on the model parameters, we describe the approximations' regime of validity and the likelihood that experiments will be in this regime, discuss the impact that the failure of these approximations would have on the predicted model, and develop theories going beyond these approximations.  Not only is it necessary to know the model parameters in order to describe experiments, but the connection that we elucidate between these parameters and the underlying assumptions that are used to derive them will allow molecule experiments to probe new physics. For example, transition state theory, which is used across chemistry and chemical physics, plays a key role in our determination of lattice parameters, thus connecting its physical assumptions to highly accurate experimental investigation.  	
\end{abstract}

\maketitle

\section{Introduction \label{sec:introduction}}

Ultracold nonreactive molecules (NRMs) have a unique interaction structure involving many interaction channels that was elucidated in Refs.~\cite{mayle:statistical_2012,mayle:scattering_2013}.  Following successes in creating ultracold chemically reactive molecules~\cite{Ni_Ospelkaus_08,PhysRevA.86.021602,PhysRevA.89.020702,deiglmayr:formation-LiCs_2008}, three ultracold dipolar NRMs have recently been produced~(RbCs~\cite{PhysRevA.85.032506,PhysRevA.89.033604,molony:creation_2014,takekoshi:ultracold-RbCs_2014,molony2016production,gregory2016controlling}, NaK~\cite{park:two-photon_2015,park:ultracold_2015,PhysRevLett.116.225306,park2016second}, and NaRb~\cite{1367-2630-17-3-035003,guo2016creation}), as have several homonuclear  species~\cite{herbig2003preparation,danzl2008quantum,danzl:ultracold-Cs2_2010,PhysRevLett.109.115303,PhysRevLett.109.115302,frisch2015ultracold,jochim2003bose}. Although the homonuclear molecules lack a dipole moment, the rich collisional physics is just as important as in the dipolar NRMs. An emerging direction is to place the NRMs in an optical lattice, which is predicted to manifest a greater variety of many-body behaviors~\cite{doi:10.1021/cr2003568,carr2009cold,lemeshko2013manipulation}.  Besides displaying a wealth of many-body phenomena, an optical lattice provides a natural means to produce NRMs at high density from a dual-species gas of ultracold atoms by Feshbach association~\cite{PhysRevLett.90.110401,PhysRevA.81.011605,PhysRevA.92.063416,moses2015creation}.

Anticipating these lattice experiments, Ref.~\cite{wall:microscopic-derivation-multichannel_2016} determined the form of the effective lattice model when ultracold NRMs are placed in a deep optical lattice. However, it provided only a formal calculation of the parameters appearing in this lattice model in terms of the solutions of a challenging four-atom problem.  Ref.~\cite{docaj:ultracold_2016} was able to predict the structure and estimate these parameters' values by relying on several approximations. 

While these approximations are realistic, they are uncontrolled and may miss some structure. Similarly, while they likely are accurate for order-of-magnitude estimates, they are not likely to be  quantitative. Gaining confidence in the structure and values of the model parameters therefore requires either more microscopic theories or experimental characterization. Quantitatively determining the parameters is not only interesting but also urgent, given the rapid pace of experimental progress in the last two years~\cite{gregory2016controlling,molony:creation_2014,takekoshi:ultracold-RbCs_2014,molony2016production,park:two-photon_2015,park:ultracold_2015,PhysRevLett.116.225306,park2016second,1367-2630-17-3-035003,guo2016creation}.  In addition to understanding complex many-body physics, elucidating the complex short-range interactions of NRMs is also essential for understanding the practical limitations of evaporative cooling schemes~\cite{zhu:evaporative_2013,stuhl2012evaporative}.

This paper presents three types of results. First, we  determine the lattice model parameters appearing in Ref.~\cite{wall:microscopic-derivation-multichannel_2016} within the framework of approximations that were used in Ref.~\cite{docaj:ultracold_2016}. Second, we describe these approximations' expected region of validity. Third, we anticipate possible deviations from the approximations and provide methods to systematically improve them.  

Experiments can test the theory that is built on the approximations presented  herein, and this  will have important consequences whether or not the theory accurately predicts measurements. If it does, the theory will establish an  effective theory to study a new regime of many-body physics. The potentially more exciting possibility is a disagreement, which will reveal that some cherished approximation has previously unknown limitations. This could have important implications even beyond ultracold physics, for example to chemistry, since the lattice model parameters connect chemical properties to many-body observables in the lattice. These properties characterize the bimolecular collisional complexes (BCCs): for example, their energies and their rates of dissociation, the central quantity in their chemical kinetics.  

An example of such an important connection to chemistry is our use of transition state theory (TST) to determine a dissociation rate of the bimolecular complexes. Given its wide use in chemistry, understanding the accuracy of TST in a controlled manner is an important goal in that field~\cite{levine:molecular_2010}. Ultracold lattice experiments could test this well-motivated, but uncontrolled, approximation with unprecedented flexibility and accuracy.  In particular, as argued in Refs.~\cite{wall:microscopic-derivation-multichannel_2016,docaj:ultracold_2016}, a deep optical lattice in which each lattice site is decoupled from the others allows for a ``chemical reaction microscope" that probes the BCCs' properties at an extremely high energy resolution -- potentially sub-nanoKelvin, even in a gas at hundreds of nanoKelvin! This energy resolution is orders of magnitude better than the already-extraordinary resolution provided
by an ultracold trapped gas in the absence of a lattice. This results from the lattice's precisely quantized energy: The fraction of molecules that are thermally excited is very small, and their contribution to the spectrum might even be relegated to sidebands that could be filtered out. Thus, there is no thermal smearing of the spectrum, and its resolution is limited only by coherence time of system.

The structure of this paper is as follows. 
Sec.~\ref{sec:overview} reviews the complexity of molecular collisions and the basic results of Refs.~\cite{docaj:ultracold_2016,wall:microscopic-derivation-multichannel_2016}. The focus is the solution of two NRMs in a harmonic well, from which the effective lattice model parameters for NRMs in an optical lattice is obtained.  Sec.~\ref{sec:approx} begins our main new results. Its subsections describe each of the approximations used in Ref.~\cite{docaj:ultracold_2016}, their regime of validity, the likelihood of the NRMs of current experimental interest being in this regime of validity, and methods to go beyond these approximations. Specifically, Sec.~\ref{sec:length-sep} focuses on the assumed separation between the three key length scales:  ``short ranges" where multi-channel interactions are important,  the van der Waals length, and  the harmonic oscillator length characterizing one site of a deep optical lattice. 
Secs.~\ref{sec:RMT}, \ref{sec:determine-wb}, \ref{sec:QDT}, and~\ref{sec:TST} focus on the approximations to the molecular interactions:  random matrix theory (RMT) used to treat the short-range chaotic motion (Sec.~\ref{sec:RMT}), methods to determine the coupling parameters appearing in the two-body Hamiltonian (Sec.~\ref{sec:determine-wb}), quantum defect theory (QDT) used to describe the effects of the van der Waals tail of the potential (Sec.~\ref{sec:QDT}), and transition state theory (TST) used to obtain the strength of coupling of the short-range BCCs to the two-molecule scattering continuum, specifically the Rice-Ramsperger-Kassel-Marcus   (RRKM) approximation (Sec.~\ref{sec:TST}).  We refer to this collection of approximations for the molecular interactions as 
RMT+QDT+TST. 
The final two subsections of Sec.~\ref{sec:approx} discuss best estimates of and the uncertainties in parameters appearing in the theory: the BCC density of states (per unit energy) and van der Waals length $\rvdw$ in Sec.~\ref{sec:molec-scatt-prop}, and the BCCs' polarizability in Sec.~\ref{sec:complex-pol}.
Sec.~\ref{sec:outlook} concludes.

\section{Overview of nonreactive molecules in an optical lattice \label{sec:overview}}
 
 \begin{figure}[t]
\includegraphics[width=0.8\columnwidth]{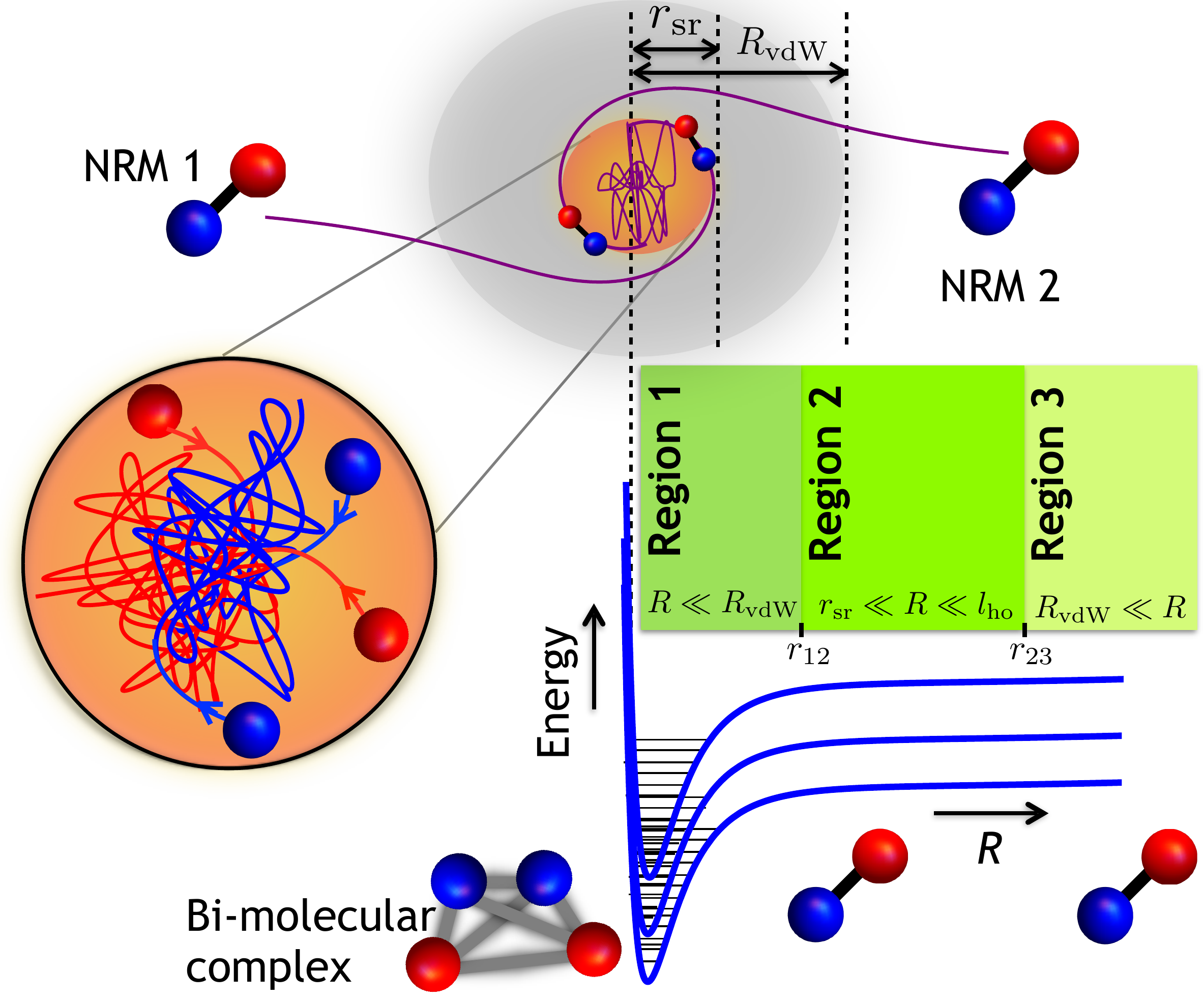}
\caption{ 
\textbf{Separation between the chaotic ($R\lsim \rsr$) and regular ($R\gsim \rsr$) regions of intermolecular scattering. }
For extreme separations $r\gg \rvdw$ the molecules propagate 
ballistically. As the molecules approach each other within the van der Waals potential but still remain in the non-chaotic, single channel region (i.e. $\rsr \lsim R\lsim \rvdw$), intermolecular interactions curve the trajectories but preserve their regularity. Finally, molecules approach within a range $R\lsim \rsr$,   and the trajectories of the constituent atoms become a chaotic tangle, mixing many internal rovibrational states. This chaotic tangle of trajectories can also be described as a superposition of bound states, the bimolecular collision complexes. 
\label{fig:chaos-ballistic-sep}
}
\end{figure}

 Reference~~\cite{wall:microscopic-derivation-multichannel_2016} derived the form of the effective lattice Hamiltonian for NRMs in a deep lattice with site occupation at most two, finding the multi-channel Hubbard model
\begin{align}
\label{eq:EffectiveModel} \hat{H}=&-J\sum_{\langle i,j\rangle, s}\left[ \hat{c}^{\dagger}_{i,s}\hat{c}_{j,s}+\mathrm{H.c.}\right]+\sum_i\left(\sum_{\alpha} U_{\alpha} \hat{n}_{i,\alpha}+\frac{3\omega}{2}\hat{n}_i\right)\,.
\end{align}
The parameters are described briefly here, with more complete definitions and discussion postponed until they are used in this paper.
This model follows from a microscopic analysis by taking advantage of the separation of length scales illustrated in Fig.~\ref{fig:chaos-ballistic-sep}. In particular, the interaction is relevant only at distances comparable to or shorter than the van der Waals length $\rvdw$. Furthermore,  $\rvdw$ is much less than the harmonic oscillator length $\lho=\sqrt{\hbar/(\mu\omega)}$ where $\mu=m/2$ is the reduced mass for two (identical) molecules of mass $m$, and $\omega$ is the angular frequency associated with the harmonic oscillator potential that approximates a single site of a deep lattice.  The single-molecule tunneling $J$ is controlled, as it is for atoms, via the optical lattice depth,  $s$ indexes the energetically available states of free NRMs (e.g., hyperfine states~\cite{Ospelkaus_Ni_10,PhysRevLett.116.225306,gregory2016controlling}, rotational excitations, or vibrational levels), and the $U_\alpha$ are determined by the eigenenergies $E_\alpha$ of the two NRMs in a harmonic oscillator. Specifically, $U_\alpha=E_\alpha-3\omega/2$ where $E_\alpha$ is the eigenenergy associated with eigenstate $\ket{\alpha}$ of the relative coordinate Hamiltonian
\be
{\hat H}_{\text{rel}} \! &=& \!\sum_n \epsilon_n \ket{n}\!\bra{n} + \sum_b \nu_b \ket{b}\!\bra{b} + \sum_{nb}\lp W_{nb}\ket{n}\!\bra{b}+\hc\rp \, , \nonumber \\
\label{eq:H-2-body-ho}
\ee
where $\epsilon_n=(2n+3/2)\omega$ and
\be
W_{nb} &=& w_b M_n/\lho^{3/2}, \label{eq:Wnb-factor}  
\ee 
with
\be
M_n &=& \sqrt{\frac{\Gamma(n+3/2)}{\Gamma(n+1)}}. \label{eq:Mn-defn} 
\ee
Here, $\lho=\sqrt{1/(\mu\omega)}$ is the harmonic oscillator length with $\mu=m/2$ the reduced mass and $m$ the mass of a single molecule, the $\ket{b}$ are short-ranged two-NRM bound states, and the $\ket{n}$ are harmonic oscillator eigenstates. 
The creation operators in Eq.~\eqref{eq:EffectiveModel} are modified from their usual form, and act on the on-site Fock states on site $i$ as 
\be 
{\hat c}_{i,s}^\dagger \ket{0}_i &=& \ket{s}_i, \\
{\hat c}^\dagger_{i,s}\ket{s'} &=& P_{s,s'} \sqrt{1+\delta_{s,s'}} \sum_\alpha \mathcal{O}_{\alpha}^{s,s'}\ket{\alpha}_i, \\
{\hat c}^\dagger_{i,s}\ket{\alpha}_i &=& 0,
\ee  
where $\ket{s}_i$ is the state with a single molecule in state $s$ in the lowest harmonic oscillator state, $\ket{s,s'}_i$ is the state with two molecules in the 
lowest energy relative harmonic oscillator eigenstate on site $i$, and 
$P_{s,s'}$ is a factor to account for fermionic exchange and Pauli blocking (see Ref.~\cite{wall:microscopic-derivation-multichannel_2016}).   The overlap factors $\mathcal{O}_{\alpha}^{s,s'}\equiv \braket{\alpha|s,s'}$
 give the weight of a particular relative coordinate eigenstate $|\alpha\rangle$ on the open-channel state $|s,s'\rangle$, and reduce the tunneling rate of an NRM onto a site containing an NRM compared to its ``bare" value $J$.

 In this paper, we determine the $\nu_b$ and $w_b$, and from these we determine
the lattice model parameters $U_\alpha$ and ${\mc O}_\alpha$, as discussed in the next section. Because accurately solving a realistic model of interacting atoms to determine the $U_\alpha$ and ${\mc O}_\alpha$ is out of present reach (see Ref.~\cite{wall:microscopic-derivation-multichannel_2016} for a formal discussion), we employ approximations.   We begin in Sec.~\ref{sec:approx} by introducing the ``standard suite of approximations" (RMT+QDT+TST) of Ref.~\cite{docaj:ultracold_2016}. We go into considerable depth into the formulation of these approximations and their regimes of validity. Then Sec.~\ref{sec:beyond-standard-approxes} discusses when these approximations may be insufficient, and describes more accurate theories that may be employed. 

\section{Approximations used to obtain parameters of the effective Hamiltonian \label{sec:approx}}

A formal derivation of the \emph{form} of Eq.~\eqref{eq:EffectiveModel}, as presented in Ref.~\cite{wall:microscopic-derivation-multichannel_2016}, requires only the separation of length scales shown in Fig.~\ref{fig:chaos-ballistic-sep}.  However, the determination of the parameters appearing in this model through microscopic means is extraordinarily difficult, and so approximate means to determine these parameters are highly desirable.  An overview of the standard RMT+QDT+TST suite of approximations to determine the model parameters was outlined in  Sec.~\ref{sec:overview}.  

Despite the multiple approximations used, the RMT+QDT+TST fit together in a fairly simple way:
\begin{itemize}
\item 
We combine approximations, 
each treating different intermolecular separations, 
using the  separation of lengths between multi- and  single-channel interactions, and the harmonic oscillator. [Sec.~\ref{sec:length-sep}]
\item RMT approximates the short distance $R\lsim \rsr$ physics where the scattering is chaotic and involves numerous interaction channels. [Sec.~\ref{sec:RMT}] 
\item QDT approximates the propagation of the molecules through the vdW potential that characterizes the large-$R$ tail of the intermolecular interactions for  $R\gg \rsr$. [Sec.~\ref{sec:QDT}]
\item TST sets the energy scale with which the short range BCCs (RMT bound states) couple to the outer region. [Sec.~\ref{sec:TST} and~\ref{sec:determine-wb}]
\item Under these approximations, 
the lattice model parameters depend on only two NRM properties, the van der Waals length $\rvdw$ and the BCCs' density of states $\rho_b$. [Sec.~\ref{sec:molec-scatt-prop} discusses the known values of these parameters and methods to estimate them.]
\end{itemize}
The relevant length scales and energies will be defined more precisely in their respective subsections.
  
The following subsections discuss the physical content 
of
each of these approximations in depth, under what assumptions they are valid, and in some detail uses them to derive the effective lattice model parameters. After the present section describes the standard suite of approximations used to obtain the model parameters, Sec.~\ref{sec:beyond-standard-approxes} will discuss possible modes of failure of these approximations, and the exciting possibility of testing the approximations in experiments.

\subsection{Separation of length scales \label{sec:length-sep}}

\subsubsection{Separation of length scales: overview and consequences}

Three length scales are crucial for NRMs in a single lattice site: $\rsr$, $\rvdw$, and $\lho$. These  are defined such that
\begin{itemize}
\item $r_{\text{sr}}$ separates short-range scattering  -- plausibly chaotic and where multiple adiabatic channels are coupled -- from  long-range scattering that consists of propagation within the open channel. A typical value (for RbCs) is roughly  $\rsr\sim 4$nm~\cite{1367-2630-12-7-073041}.
\item $\rvdw$ is the characteristic length scale associated to the van der Waals potential $V(R)= -C_6/R^6$; this is $\rvdw=(2\mu C_6)^{1/4}$. A typical value (for RbCs) is roughly $\rvdw\sim25$nm~\cite{1367-2630-12-7-073041,julienne:universal_2011}. 
\item  $\lho=\sqrt{1/(\mu\omega)}$ is the characteristic length scale associated with the harmonic trap.  Typical values for optical lattices where both tunneling and interactions are relevant are $\lho\gsim 100$nm.  (This $\lho$ would arise for typical numbers $\mu\sim 100\text{amu}$ and $\omega\sim 2\pi \times 10 \text{kHz}$, where amu is the atomic mass unit.) 
\end{itemize}
The methods used to obtain these values, their dependence on molecular species and other parameters, and their uncertainties are explained in Sec.~\ref{sec:molec-scatt-prop}.  

The derivation in Ref.~\cite{docaj:ultracold_2016} and Sec.~\ref{sec:overview} assumed the separation of lengths $\rsr\ll \rvdw\ll \lho$, as illustrated in Fig.~\ref{fig:chaos-ballistic-sep}.
There is little difference in our assumption of this separation of lengths for NRMs and the corresponding assumptions that are made for ultracold atoms in an optical lattice that lead to the Hubbard model. For atoms, the separation between the lengths 
$\{\rsr,\rvdw\}\ll \lho$ is what enables one to use a pseudopotential parameterized by a scattering length to describe the  interatomic interactions in a lattice and derive the conventional Hubbard $U$~\cite{jaksch_bruder_98}. For NRMs, the primary difference with atoms is not due to any difference in this separation of scales, but rather the complex, energy-dependent interaction physics occurring for $R\lsim \rsr$.

To understand how these separations appear in and impact the calculation, we consider three regions of intermolecular separation, depicted in Fig.~\ref{fig:chaos-ballistic-sep}, and summarized as follows:  
\begin{itemize}
\item \textbf{Region 1}: intermolecular separations $R$ such that $R<r_{12}$. We will choose $r_{\text{12}}$ such that in this region $R \ll \rvdw$. This enables two approximations: One can treat the dynamics as chaotic, due to the complexity of the potential in this region, and one can ignore the effects of the harmonic oscillator potential since it is constant over this region. 
\item \textbf{Region 2}: $r_{12}<R<r_{23}$. We will choose $r_{12}$ and $r_{23}$ such that $r_{\text{sr}} \ll R \ll \lho$ in this region, so that one can solve for the wavefunction in a single channel vdW potential, ignoring the short range physics and treating the trap potential as a constant.  
\item \textbf{Region 3}: $R>r_{\text{23}}$. We will choose $r_{\text{23}}$ such that  $R\gg \rvdw$ in this region (and therefore $R\gg \rsr$ as well). Here, one can calculate the wavefunctions in the presence of the trap potential only. 
\end{itemize}
Importantly, one can choose  the $r_{12}$ and $r_{23}$ such that the requisite inequalities in each region are simultaneously satisfied. To create the regions in such a manner, we choose $r_{\text{12}}=\sqrt{r_{\text{sr}}\rvdw}$ and $r_{\text{23}}= \sqrt{\rvdw \lho}$.

In \textbf{Region 1}, the separation $R\ll \{\rvdw,\lho\}$ allows us to calculate the wavefunctions by taking the dynamics to be chaotic, and it allows us to ignore the trap potential, since it is
 effectively constant over this region. Moreover, it allows us to treat this region ``classically" in the sense that we can ignore the near-threshold effects (e.g. Wigner laws) arising from the large-$r$ vdW tail. These become important when the (relative coordinate) kinetic energy gives a de Broglie wavelength comparable to the length scale on which the potential varies (e.g. $\rvdw$). However, at short-range, the potential is so deep that  the kinetic energy is large and  the de Broglie wavelength is small compared to $\rvdw$.  This classicality is a necessary requirement to apply the TST to calculate $\nu_b$s and $w_b$s (see Sec.~\ref{sec:TST}).

In \textbf{Region 2}, the separation $R\gg \rsr$ ensures that one can calculate the eigenstates including only the open channel potential and ignoring the interchannel couplings. The separation $R \ll \lho$ allows one to approximate the open channel potential as a $-C_6/R^6$ potential. Consequently, the wavefunction in this region is the solution to a single-channel problem in the vdW potential. This wavefunction can be calculated using QDT, which is designed to solve such a problem with arbitrary boundary conditions (see Sec.~\ref{sec:QDT}).

In \textbf{Region 3}, the separation $R \gg \rvdw$ makes the  basis of harmonic oscillator eigenstates $\ket{n}$ natural. In the absence of the short-range interactions at $R\lsim \rvdw$, the $\ket{n}$ would be eigenstates  and the Hamiltonian would be diagonal in this basis. The short-range interactions  introduce matrix elements between these eigenstates and closed channel bound states $\ket{b}$ (and in principle between the $\ket{n}$), but these matrix elements may be taken to be confined near the origin at a spatial scale much less than $\lho$. Therefore, the harmonic oscillator eigenstates may be treated as constant over the length scale on which the interaction potential varies. This separation was employed in Sec.~\ref{sec:overview} to factor $W_{nb}=w_b M_n/\lho^{3/2}$ in Eq.~\eqref{eq:Wnb-factor}, which is derived in detail in Ref.~\cite{wall:microscopic-derivation-multichannel_2016}.

\begin{figure*}
\includegraphics[width=1.9\columnwidth]{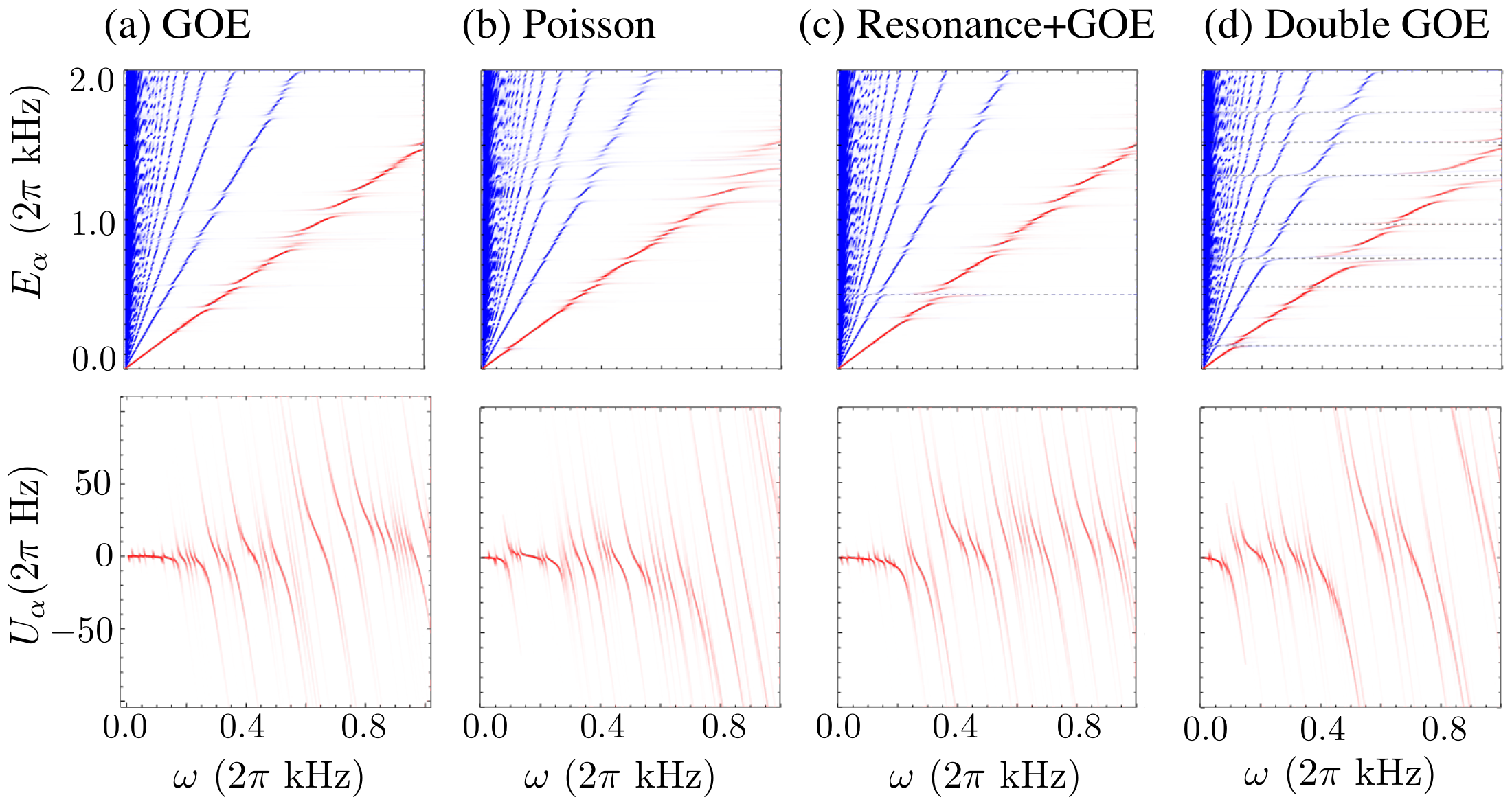}
\caption{\textbf{Eigenvalues $E_\alpha$ (top row) and effective interactions $U_\alpha$ (bottom row) as a function of trap frequency $\omega$.} Darkness (i.e. opacity) is set proportional to the open-channel weight ${\mc O}_\alpha$, indicating the importance of the corresponding state to the lattice model physics. 
(a)  Gaussian Orthogonal Ensemble (GOE) for parameters corresponding roughly to RbCs [eigenvalues are from Hamiltonians sampled from Eq.~\eqref{eq:sr-ham-prob-eq}]. 
(b) Independent levels (Poisson statistics) [Eq.~\eqref{eq:poisson-nu-b}]. 
(c) Broad resonance on GOE background. 
(d) Two-ensemble model [Eq.~\eqref{eqs:2-RMT}].  Parameters used to generate these plots are discussed in the main text.
\label{fig:beyond-RMT-model-props}}
\end{figure*}

Consequences of the separation of lengths are apparent in Fig.~\ref{fig:beyond-RMT-model-props}(a), which shows the results for the eigenenergies of two NRMs on one lattice site,  $E_\alpha$, and the associated interaction parameters $U_\alpha$ calculated within the suite of approximations that we present here. One consequence of the separation of lengths is  that the bound state energies, which hybridize with the oscillator states, are $\omega$-independent. A second consequence is that the energy splittings of the avoided crossings and the typical $U_\alpha$ acquire the characteristic $n$- and $\omega$-dependences of $W_{nb}$. In particular, the panel shows demonstrates that width of the resonances and typical size of the $U_\alpha$s clearly increases with increasing $\omega$ and with increasing $n$, as expected from $W_{nb}$'s dependence on $n$ and $\lho$.  The figures were generated with parameters $R_{\text{vdW}}=25$nm, a reduced mass of $\mu=110$amu, and $\rho_b=1/(2\pi \times 20 \text{Hz})$. To numerically sample the GOE we used 500 bound states and we included 100 harmonic oscillator states. 
The remaining panels go beyond this standard suite of approximations and are discussed in detail in Sec.~\ref{sec:beyond-standard-approxes}.

\subsection{Random matrix theory for bimolecular complexes (bound states) \label{sec:RMT}}

As Fig.~\ref{fig:chaos-ballistic-sep} suggests, at short ranges $R\lsim \rsr$, the dynamics is chaotic and involves many interaction channels. While this situation may appear extremely complicated, it suggests taking advantage of the ability of chaos to smear out detailed structure in dynamics. The tool we use to do this is RMT applied at these short ranges, in particular to the bound state energies $\nu_b$ and their couplings to longer range states $w_b$.
This section discusses the physical origin of the RMT and
 when this approximation is
expected to be valid. 

\subsubsection{Physical origin of RMT: which Hamiltonian is being treated as a random matrix?}
At the 
broadest level, RMT is expected to describe the statistical properties of generic
 observables in quantum mechanical systems when the classical dynamics of the system is chaotic (although this has not been proven rigorously). This conjecture is sometimes referred to as the Bohigas-Giannoni-Schmit conjecture~\cite{PhysRevLett.52.1}. 
 One formulation of the key statement of RMT is that the 
Hamiltonian of the system is a random matrix sampled from some probability 
distribution.  

It is plausible that chaotic dynamics and therefore RMT will somehow manifest in 
ultracold molecular scattering. Many examples of chaotic scattering between particles in physics are known, 
spanning from complex ultracold atoms at nanoKelvin temperatures~\cite{frisch2014quantum,maier2015broad,maier2015emergence} to
interactions within the nucleus at energies of $\sim \! 10\text{MeV} \sim 10^{11}$K~\cite{mitchell:random_2010}. Remarkably this unifies the physics of these collisions occurring at energies separated by nearly twenty orders of magnitude! Molecular scattering is
usually expected to be chaotic for complex molecules, but it is less clear how chaotic the scattering of simple diatomic molecules is. Nevertheless, the complexity of the interatomic potentials and significant number of particles involved (four atoms, many more electrons) plausibly will lead to chaos. This suggestive observation has 
been corroborated by observation of chaotic dynamics in classical molecular dynamics simulations of diatomic molecule scattering using model potentials~\cite{PhysRevA.89.012714}, as well through level spacing analysis in other complex systems, such as atom-molecule~\cite{PhysRevA.93.052713} and alkaline-earth collisions~\cite{PhysRevA.93.022703}.

Despite the chaotic scattering, at first blush it is unclear that a random matrix approach will be useful in this situation: The dynamics is obviously not fully chaotic since once the particles have scattered and are well outside the range of the intermolecular interaction potential, they simply propagate ballistically. Therefore, we need to consider carefully 
exactly what Hamiltonian is being treated as a random matrix and how it combines 
with the other, non-chaotic, parts of the Hamiltonian.

To apply RMT we note the  separation of scales of the scattering problem between small-separation chaotic dynamics and large-separation regular (and eventually ballistic) dynamics, which is illustrated in Fig.~\ref{fig:chaos-ballistic-sep}. At short distances  $R\lsim \rsr$, the classical dynamics indeed appears chaotic~\cite{PhysRevA.89.012714}. Furthermore, this short-range regime allows strong coupling between the interaction channels. Outside this, the interactions are much simpler (single channel van der Waals or simply negligible). Therefore, it is useful to decompose the full Hilbert space of four atoms (two molecules) $\mc H_{2m}$ into a direct sum of two terms: ${\mc H}_{2m}={\mc H}_{\text{b}}+{\mc H}_{lr}$. Here ${\mc H}_{\text{b}}$ is a ``short range Hilbert space" consisting of the four-atom wavefunctions with support on configurations where all four atoms are within $\rsr$ of each other\footnote{The notation ${\mc H}_{\text{b}}$ is to hint that this is where the bound states live, as will become central later.} (there is some freedom in this choice).  Conversely, ${\mc H}_{\text{lr}}$ is the Hilbert space of wavefunctions with support on configurations where at least two atoms are more than $\rsr$ apart.

We can associate a Hamiltonian with each of  these Hilbert spaces. The total relative coordinate
 Hamiltonian is 
\be
{\hat H}_{\text{rel}} &=& \Hb +{\hat H}_{\text{lr}}+{\hat H}_{\text{cpl}}, \label{eq:separated-hams}
\ee
where $\Hb$ is ${\hat H}_{\text{rel}}$ projected onto ${\mc H}_{\text{b}}$, ${\hat H}_{\text{lr}}$ is ${\hat H}_{\text{rel}}$ projected onto ${\mc H}_{\text{lr}}$, and ${\hat H}_{\text{cpl}}$ is the remaining part of the 
Hamiltonian, which couples the short- and long-ranged Hilbert spaces. 
Choosing a basis $\ket{b}$ of eigenstates of $\Hb$ and a basis $\ket{n}$ of eigenstates of $\hat{H}_{\text{lr}}$, the three terms in this equation map onto the three terms in Eq.~\eqref{eq:H-2-body-ho}. Note that we are not assuming that  Eq.~\eqref{eq:H-2-body-ho} holds; rather, we are starting from a $\hat{H}_{\text{rel}}$ for which only general properties are known, and deriving Eq.~\eqref{eq:H-2-body-ho} and the form of its couplings from it. A microscopic description of such an $\hat{H}_{\text{rel}}$ was provided in Ref.~\cite{wall:microscopic-derivation-multichannel_2016}.

Under the assumption that $\Hb$ describes fully chaotic dynamics,  its statistical
properties are expected to be well-described by RMT~\cite{mitchell:random_2010,RevModPhys.53.385,mehta2004random}. 
RMT states that $\Hb$ can be taken to be sampled from a 
probability distribution 
\be
P(\Hb) &=& {\mc N}_H e^{-\operatorname{Tr} {\Hb}^2/2\sigma^2}\, , \label{eq:sr-ham-prob-eq} 
\ee 
with 
${\mc N}_H$ ensuring normalization of the probability distribution. 
The GOE distribution is chosen because it is the random matrix ensemble that describes models with time-reversal symmetry. Even if the experiments are performed in a magnetic field, for any realistic values in an ultracold experiment this field is too weak to strongly mix scattering channels, so the GOE remains appropriate. 
This result together with Eq.~\eqref{eq:separated-hams} (and the structure of the Hilbert spaces on which the Hamiltonians in that equation were defined to act) allows us to determine the $\nu_b$s and the $w_b$s, up to an overall scale for the $w_b$ (which will be determined from TST in Sec.~\ref{sec:TST}.) 

\subsubsection{Determining model parameters from RMT}

\textbf{Determining $\nu_b$s.} 
Since the $\nu_b$s are defined -- by Eq.~\eqref{eq:H-2-body-ho} -- as the eigenvalues of $\Hb$, we sample $\Hb$ according to Eq.~\eqref{eq:sr-ham-prob-eq} and solve for its $N_b$ eigenvalues $\nu_b$.  
One finds that the average density of eigenvalues approaches a semicircle on $(-\sqrt{2N_b }{\sigma},\sqrt{2N_b }{\sigma})$, namely
$\rho_b^{\text{sc}}(\nu)=\frac{1}{\pi \sigma^2} \sqrt{2N_b \sigma^2-\nu^2}$,
the so-called Wigner semicircle distribution~\cite{mehta2004random}. We wish to mimic a distribution of eigenvalues of the NRM bound states that is uniform over the energy range that is relevant to the lattice physics. Thus we choose a large enough semicircle (large enough $N_b$) such that the distribution is roughly constant over the relevant scale. At the peak the Wigner semicircle density of states is equal to $\rho_b^{\text{sc}}(0)=\sqrt{2N_b}/(\pi \sigma)$, so we choose 
\be 
\sigma &=& \frac{\sqrt{2N_b}}{\pi\rho_b} \label{eq:RMT-sigma-eq}
\ee
to ensure the proper density of bound states near zero energy.

\begin{figure}
\includegraphics[width=0.7 \columnwidth]{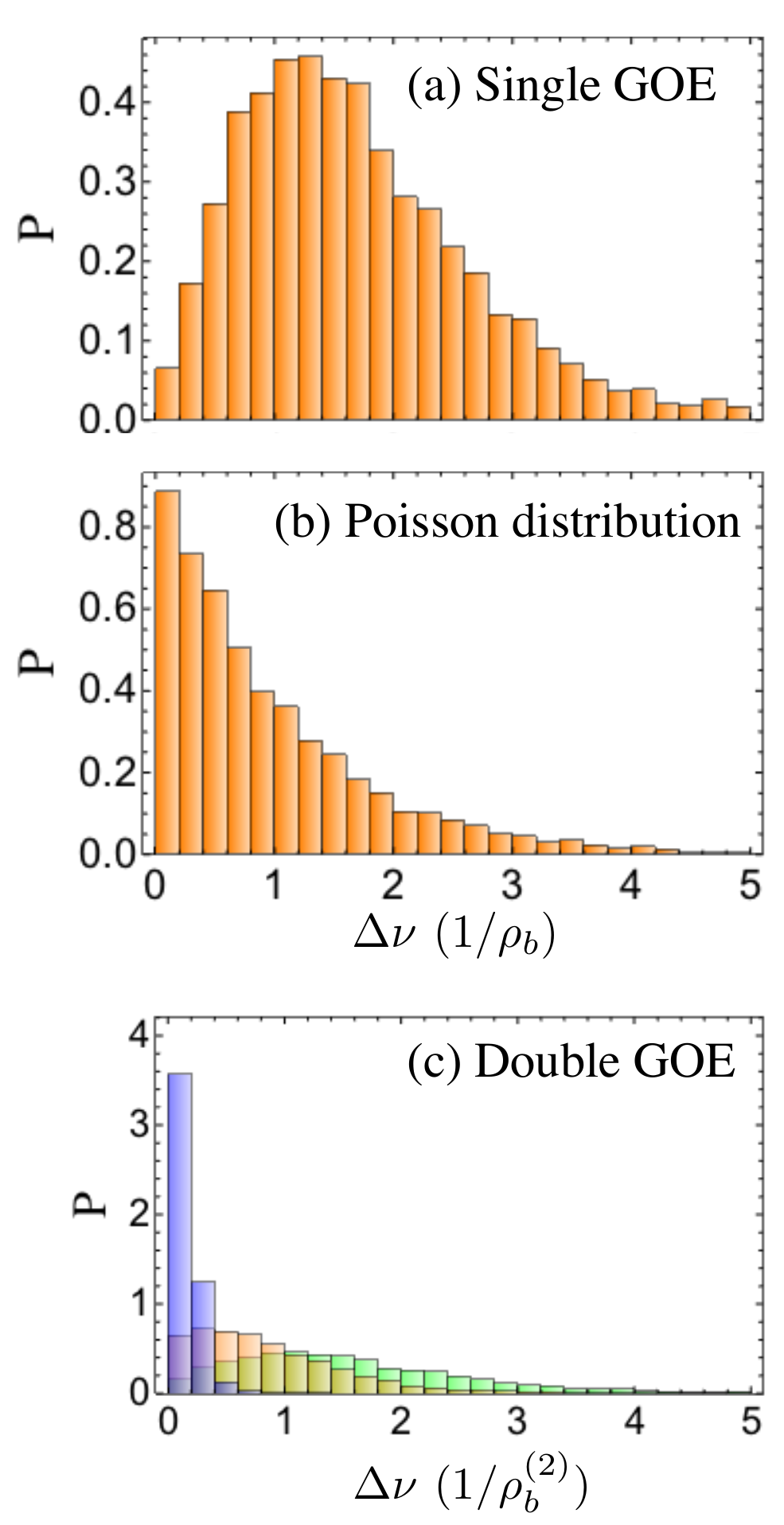}
\caption{\textbf{Probability distribution of nearest-level spacings $\Delta$ in different
 matrix ensembles.}   (a) Random matrix theory Gaussian Orthogonal Ensemble (GOE) [eigenvalues are from Hamiltonians sampled from Eq.~\eqref{eq:sr-ham-prob-eq}]. 
(b) Independent levels (Poisson statistics) [Eq.~\eqref{eq:poisson-nu-b}]. 
(c) Two-ensemble model [Eq.~\eqref{eqs:2-RMT}] for $\rho_b^{(1)}/\rho_b^{(2)}=0.1, 1.0, 10.0$ from ``left to right" (blue, orange, and green, respectively). 
 \label{fig:nearest-level-prob-dist} }
\end{figure}

Another interesting characteristic of the eigenvalues is the level spacing distribution. This is the probability distribution $p(\Delta \nu)$ of the difference $\Delta \nu^{(j)} =\nu_b^{(j)}-\nu_b^{(j-1)}$, where $j$ indexes the eigenvalues from lowest to highest, with the distribution taken over the ensemble of all $j$ and all $\hat{H}_{\mathrm{b}}$ via Eq.~\eqref{eq:sr-ham-prob-eq}. This distribution is shown in   Fig.~\ref{fig:nearest-level-prob-dist}(a) and is well-approximated by the ``Wigner surmise" $p(\Delta \nu)=(\pi/2\lambda^2) \Delta \nu \,e^{-(\pi/4)(\Delta \nu/\lambda)^2}$ where $\lambda$ is the average level spacing.  We note that higher order correlations between the levels (three level distributions, etc.) are also nontrivial~\cite{mehta2004random}.

\textbf{Determining $w_b$s.} Obtaining the statistical distribution of the $w_b$s requires a little more effort. Note that the $w_b$s are given in terms of the matrix elements $W_{nb}=\braket{b|{\hat H_{\text{rel}}}|n}$ of Eq.~\eqref{eq:H-2-body-ho} by
\be 
w_b&=& \frac{\lho^{3/2}}{M_n} W_{nb}.  \label{eq:wb-Wnb-rel}
\ee
We make further progress by realizing that the $W_{nb}$s are matrix elements involving $\ket{b}$'s, states that we are taking to be the eigenvectors of the Hamiltonian $\Hb$ that we are sampling with the RMT. In particular, note that $W_{nb}$ is the matrix element between this random vector $\ket{b}$ and the vector $\bra{v}\equiv \bra{n}\!{\hat H}_{\text{rel}}$, i.e. 
\be 
W_{nb}&=& \braket{v|b}.
\ee
To determine $W_{nb}$'s distribution, the task at hand is: Given a vector $\ket{v}$ (not necessarily normalized), determine the distribution of overlaps of the random eigenvectors $\ket{b}$ with $\ket{v}$. To do this, it is helpful to write $\ket{v}=N_v \ket{\tilde v}$ where $N_v=\sqrt{\braket{v|v}}$ (so $\ket{\tilde v}$ is the normalized state). Then one has 
\be 
W_{nb} &=& N_v\braket{{\tilde v}|b}.\label{eq:Wnb-normalized-v}
\ee
 We can write the vector $\ket{b}$ as  
\be 
\ket{b} &=& b_1 \ket{\tilde v} + b_2 \ket{v_2} + \cdots + b_M \ket{v_M}
\ee 
where we have chosen an orthonormal basis consisting of $\ket{\tilde v}$ and $M-1$ additional vectors $\ket{v_{j=2,\ldots, M}}$ for the short-range Hilbert space $\Hb$. Although this may be an infinite dimensional Hilbert space, we handle this by considering a finite basis of dimension $M$ and let $M\rightarrow \infty$. Now we determine the distribution of $\braket{b|\tilde v}=b_1$.

Since the probability distribution of $\Hb$ is invariant under orthogonal transformations, so must be the probability distribution of eigenstates $\ket{b}$. Consequently, we are looking for the probability distribution of unit-magnitude $\ket{b}$s that is invariant under  orthogonal transformations. Consider the probability distribution  
\be 
P_b(b_1,b_2,\ldots,b_M) = P(\v{b})&=& {\mc N}_b e^{-\sum_j |b_j|^2/(2\xi^2)} \label{eq:eigenstate-prob-dist}
\ee
for the components of $\ket{b}$, where ${\mc N}_b$ is a normalizing factor and $\xi$ is a to-be-determined constant.
This distribution is invariant under orthogonal transformations  since it depends only on the manifestly invariant inner product of $\vec{b}$ with itself. To determine $\xi$, we enforce the condition that the vectors $\v{b}$ are normalized. One necessary condition for the normalization is that it holds on average: the average normalization of the eigenstates should be unity. Denoting averages over the ensemble $P_b(\vec{b})$ with overbars, i.e. $\overline{\cdots}$,  requiring normalization on average implies $\overline{b^2}=1$, so
\be 
1 &=& \sum_j \overline{b_j^2}= M \xi^2\, ,
\ee
and therefore
\be 
\xi &=& 1/\sqrt{M}.
\ee
Although this ensures that the eigenstates are normalized on average, it doesn't ensure that each eigenstate is normalized. On the contrary, it worryingly appears that the normalization  fluctuates since it is the sum of $M$ random numbers. However, the fluctuations of the norm $\overline{(b^2)^2} - (\overline{b^2})^2$ are $O(1/\sqrt{M})$, and are thus negligible for $M\rightarrow \infty$.
This shows that the ensemble of $\ket{b}$s in Eq.~\eqref{eq:eigenstate-prob-dist} gives the probability distribution of normalized vectors that is invariant under orthogonal transformations, at least as $M\rightarrow \infty$.

The distribution Eq.~\eqref{eq:eigenstate-prob-dist} of $b_1=\braket{\tilde v|b}$ and Eq.~\eqref{eq:Wnb-normalized-v} implies that
$P_W(W_{nb}) = {\mc N}_W e^{-|W_{nb}|^2/2(N_v \xi)^2 },
$. Consequently, Eq.~\eqref{eq:wb-Wnb-rel} gives
\be 
P_w(w_b) &=& \mc N_w e^{-w_b^2/(2\sigma_w^2)} \label{eq:wb-gaussian}
\ee 
with $\sigma_w = \frac{\lho^{3/2}}{M_n} \frac{N_v}{\sqrt{M}}$.  Note that from the separation of lengths discussion (see 
Secs.~\ref{sec:overview}, \ref{sec:length-sep} for more details, as well as Ref.~\cite{wall:microscopic-derivation-multichannel_2016} for a  microscopic viewpoint), we expect $N_v \propto M_n/\lho^{3/2}$ so that $\sigma_w$ does not depend on our choice of $n$ for this derivation and is independent of $\lho$. 
This calculation does not determine $\sigma_w$ since we do not know $N_v/\sqrt{M}$. (Note that $N_v/\sqrt{M}$ is finite even as $M\rightarrow\infty$ due to the $M$ dependence of $N_v$.) The parameter $\sigma_w$ will be determined by a combination of QDT+TST in Secs.~\ref{sec:determine-wb}, \ref{sec:QDT} and~\ref{sec:TST}. 

In  the continuum limit ($\omega\rightarrow 0$) this method of determining the short-range Hamiltonian and its couplings to the long range sector of the Hilbert space is equivalent to that used to obtain the short-range scattering (via the $K$-matrix) in Refs.~\cite{mayle:statistical_2012,mayle:scattering_2013}. Ref.~\cite{mitchell:random_2010} provides a review in other contexts. Although the Hamiltonian formulation is equivalent to the scattering theory, the former is convenient for many purposes. For example, it allows for straightforward building of the Hamiltonian as a matrix in a basis and calculating its eigenvectors.

Figure~\ref{fig:beyond-RMT-model-props}(a)'s lattice model parameters ${\mc O}_\alpha$ and $U_\alpha$ reflect the RMT distribution of $\nu_b$ and $w_{b}$. The random distribution of resonances is the reflection of the random distribution of $\nu_b$s, and the distribution of $U_\alpha$ is due to the random distribution of $w_b$s. 
The ${\mc O}_\alpha$s are largest when the bound states come within $w_b$ of the open channel harmonic oscillator states, creating hybridized eigenstates with significant weight on the open channel harmonic oscillator state.

\subsection{Determining $\sigma_w$: QDT+TST \label{sec:determine-wb}}

\begin{figure*}
\includegraphics[width=1.25\columnwidth]{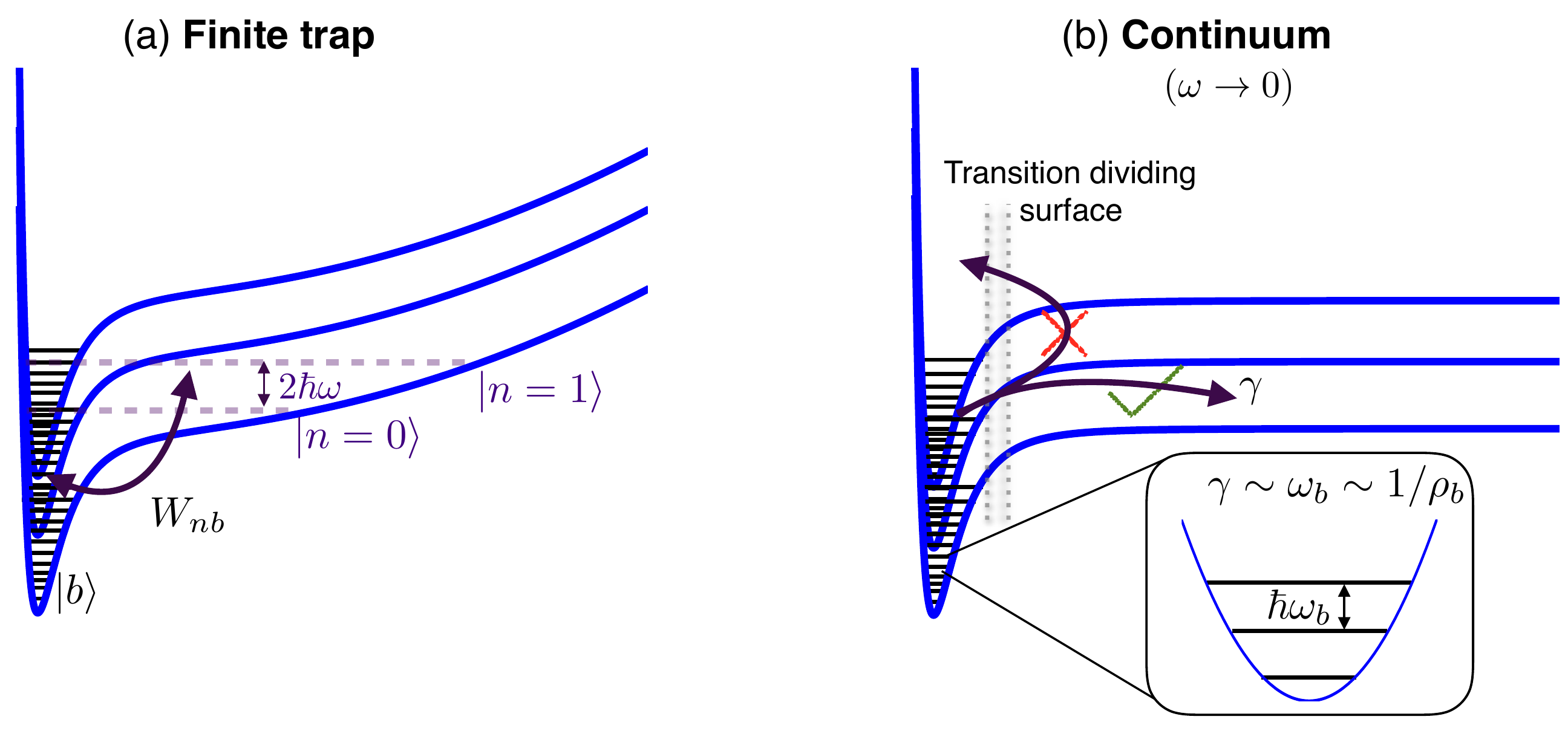} 
\caption{\textbf{Determining $W_{nb}$'s 
from TST.} (a) The $W_{nb}$ are the coupling between the closed channel bound state $\ket{b}$ and the open channel harmonic oscillator states $\ket{n}$. (b) $W_{nb}$ governs the  dissociation rate $\gamma_b$ of the BCC (bound state) $\ket{b}$ breaking into two molecules in the absence of a trap ($\omega\rightarrow 0$). By approximating $\gamma_b$ within TST+QDT, we calculate the $W_{nb}$. (c) The essence of TST for the present barrierless reactions is that the bound state oscillates at frequency $\omega_{cc} \sim 1/\rho_b$,  passes through the transition dividing surface (roughly where the vdW potential becomes negligible) with $O(1)$ probability, and then never recrosses this surface.
\label{fig:TST-cartoon}}
\end{figure*}

This section describes the method that we have used to obtain the scale of the 
couplings $w_b$, i.e. $\sigma_w$ in Eq.~\eqref{eq:wb-gaussian}.  
The structure of these couplings -- that they are drawn from
the Gaussian distribution -- follows from the RMT as described in 
Sec.~\ref{sec:RMT}, but the standard deviation of that distribution, $\sigma_w$,
is undetermined from RMT. In principle, $\sigma_w$ could be determined by 
experimentally measuring the $U_\alpha$ and ${\mc O}_\alpha$ in the Hamiltonian 
Eq.~\eqref{eq:EffectiveModel} and choosing the $\sigma_w$ that reproduces their statistics. 
These experiments remain to be done, and even after they are it will be valuable 
to have expectations for the scale of $\sigma_w$ and its dependences on parameters
such as the lattice depth and molecular species.

Although one approach is to calculate the full four-atom eigenstates from the microscopic Hamiltonian including the relevant interatomic interactions, 
in practice solving these equations is beyond the reach of current numerical methods. 
In fact, even obtaining the interatomic interaction potential appearing in the equations to sufficient accuracy is challenging.  Nevertheless, Ref.~\cite{wall:microscopic-derivation-multichannel_2016} formally shows how the couplings may be obtained from the solutions to the coupled-channel Schr{\"o}dinger equation in principle, if one were able to solve for its eigenstates. This is illuminating since it sheds light on what the couplings are microscopically, and advances in numerical algorithms may allow the problem to be solved in the future. 

To determine the $\sigma_w$ we will relate $w_b$ to the dissociation rate $\gamma_b$ of bound state $\ket{b}$. This inverts our usual picture: rather than thinking of scattering -- where two incoming molecules in a scattering state couple to the BCC (bound state) and then exits through an outgoing scattering state -- we start with the NRM in the bound state and consider its dissociation. As Fig.~\ref{fig:TST-cartoon} illustrates, both of these processes are determined by the $w_b$; adopting the latter perspective will make it easier to connect the $w_b$ to physical properties.

To obtain $\gamma_b$, we will rely on approximations developed in chemistry (TST) and low-energy scattering (QDT). This approach is illustrated in Fig.~\ref{fig:TST-cartoon}. Imagine a molecule initially in the BCC $\ket{b}$ in free space (no trap or lattice).
 It decays with a rate $\gamma_b$ into the continuum of two-molecule states (where the two molecules are far apart). Applying Fermi's Golden Rule to ${\hat H}_{\text{rel}}$
for two molecules in a lattice site, Eq.~\eqref{eq:H-2-body-ho}, taking $\ket{b}$  to be the initial state and 
$W_{nb}$ to be the perturbation, gives  
\be
\gamma_b &=& 2\pi \frac{w_b^2}{\lho^3}\sum_n M_n^2 \delta(\nu_b-\epsilon_n). \label{eq:decay-from-effective-ham}
\ee
We consider the continuum (no-lattice) limit of this ($\omega\rightarrow 0$ while fixing the bound state energy $\nu_b$), finding
\be
\gamma_b &=& \frac{\pi \mu^{3/2}\sqrt{\nu_b} w_b^2}{\sqrt{2}}
\hspace{0.5in} \text{as $\omega\to 0$}. \label{eq:decay-from-effective-ham-continuum} 
\ee
 We focus on the continuum limit because the approximations TST+QDT that we will apply later to calculate $\gamma_b$ apply to the  dissociation of bound states into the continuum.  
One can solve this equation for $w_b$ given $\gamma_b$. Actually, we will only be able to obtain the \textit{average} dissociation rate $\gamma_b$, but it turns out that within the RMT this is all we need. The average dissociation rate (which we will still call $\gamma_b$)
\be 
\gamma_b &=& \frac{\pi \mu^{3/2}\sqrt{\nu_b} \sigma_w^2}{\sqrt{2}} \label{eq:decay-from-effective-ham-continuum-2}
\ee
since $\overline{w_b^2}=\overline{w_b^2}-(\overline{w_b})^2=\sigma_w^2$, denoting RMT averages with $\overline{\cdots}$. This allows us to solve for $\sigma_w$ as
\be
\sigma_w &=& \lp\frac{2}{\mu^3 \nu_b}\rp^{1/4} \sqrt{\frac{\gamma_b}{\pi }}. \label{eq:sigmaw-from-gammab}
\ee 
Once we determine $\gamma_b$, this will give us $\sigma_w$. 

To determine $\gamma_b$, we turn to the TST+QDT combination of approximations. The TST+QDT approximations relate $\gamma_b$ to observable molecular properties, in particular $\rho_b$ and $\rvdw$. The TST determines the rate $\gamma_{\text{TST}}$ for the bound state to dissociate at short-range, e.g., to leave the radius $r_{\text{sr}}$, while the QDT determines the probability $A(\nu_b)$ for a pair of NRMs that have dissociated at $r_{\text{sr}}$ to propagate out to $R\gg \rvdw$.  The 
decay rate
$\gamma_b$ from BCC to two-molecule scattering states is obtained by stitching these two
approximations together, which yields  (see Sec.~\ref{sec:QDT})
\be 
\gamma_b &=& \gamma_{\text{TST}} {\mc A}(\nu_b). \label{eq:decay-from-TST+QDT}
\ee
We discuss the QDT and TST factors in  
Sec.~\ref{sec:QDT} and Sec.~\ref{sec:TST}, respectively.  Using those results to obtain $\gamma_b$ via Eq.~\eqref{eq:decay-from-TST+QDT}, we will soon see that using this $\gamma_b$ in Eq.~\eqref{eq:sigmaw-from-gammab}determines $\sigma_w$ in terms of molecular parameters, given by Eq.~\eqref{eq:sigmaw-final}.

\subsection{Quantum Defect Theory \label{sec:QDT}}
 
In this section we calculate the propagation in the vdW potential that gives the parameter ${\mc A}$ in Eq.~\eqref{eq:decay-from-TST+QDT}. To determine this requires solving the single-channel problem in a vdW potential at low energy. We do this, obtaining analytic expressions, using the framework of quantum defect theory. 

At the short intermolecular separations where the BCCs are bound, typical kinetic and potential energies are $\sim10^{3}$K, many orders of magnitude larger than the $10^{-3}$K energy scales associated with the excitations of individual, well-separated molecules.  However, physics at long-range is governed by a fundamentally different set of channels (e.g.~hyperfine levels) which are sensitive to externally applied fields and threshold effects from low collision energy.  Multi-channel quantum defect theory (MQDT) leverages this vast separation of length and energy scales in scattering problems by finding a representation of wavefunctions in the long-range tail of the potential that depends only weakly on energy, and matching these wavefunctions to short-range, strongly coupled physics at a short-range ``matching radius" $\rsr$~\cite{PhysRevLett.81.3355,ruzic:quantum_2013}.  

\begin{figure}[t]
\includegraphics[width=.97\columnwidth]{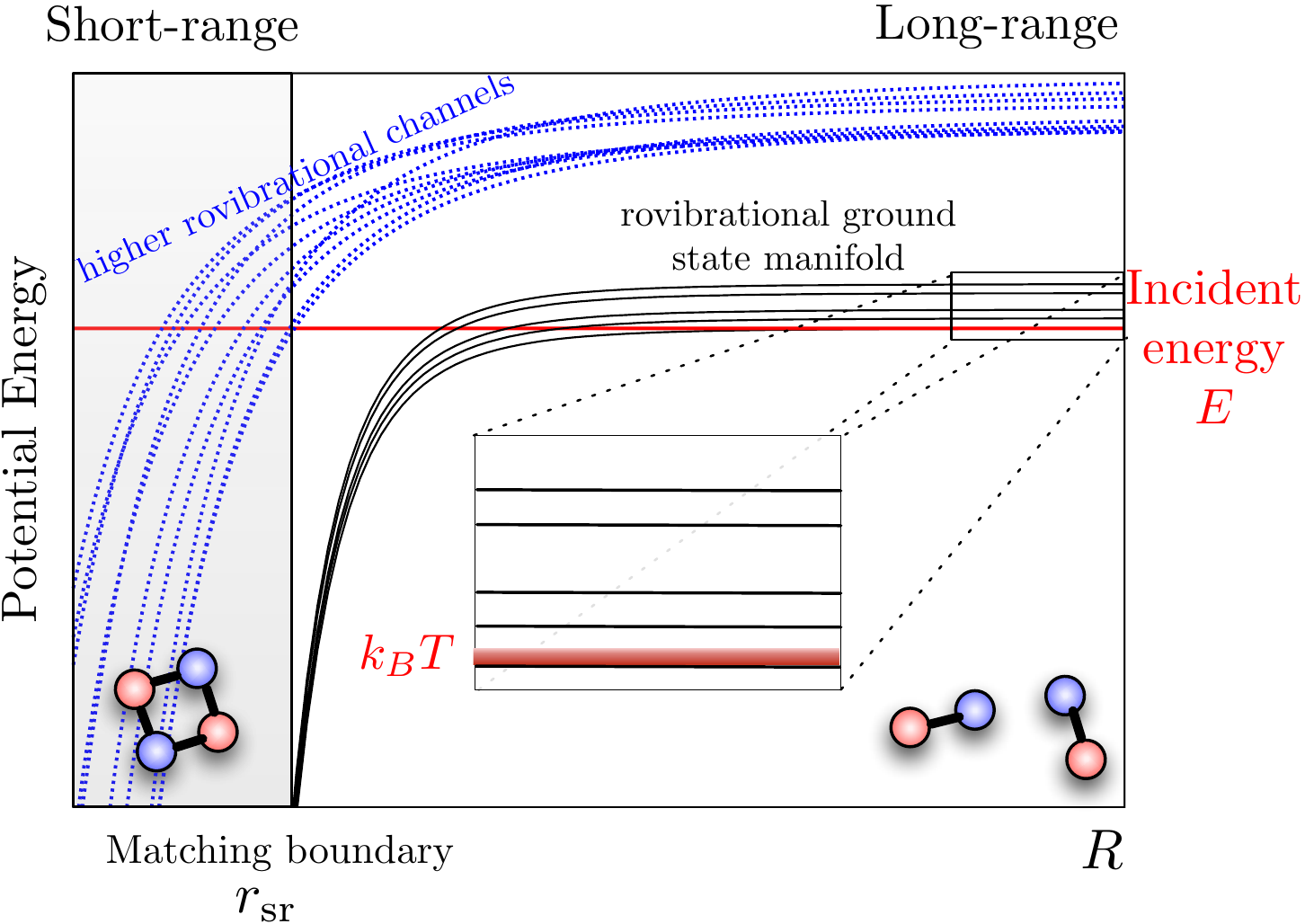}
\caption{(color online) \textbf{Schematic of multi-channel quantum defect theory.}  In multi-channel quantum defect theory (MQDT), the separation coordinate space is divided into short-range and long-range, parameterized with respect to $\rsr$.  A set of reference wave functions for the manifold of low-energy states (black solid lines) are constructed to minimize the energy dependence of the coupling to short range, and account for propagation in the $1/R^6$ van der Waals tail.  Inset shows that the separation between states in the lowest manifold is generally much larger than the temperature.  The cutoff $\rsr$ is chosen so that the states in the lowest manifold which are closed as $R\to \infty$ are classically open at $\rsr$.  For NRMs, where many channels (blue dashed lines) remain closed at $\rsr$, these channels impart a complex resonance structure to the short-range physics.  For polar NRMs, these higher-lying channels also have a qualitatively different $1/R^3$ long-range potential character.
\label{fig:MQDT}}
\end{figure}

In our particular case, the interaction potential's matrix elements (between various channels) at large separation is dominated by long-range interactions of the form
\begin{align}
\nonumber  \langle c|V\left(R\right)|c'\rangle=&\left[-\frac{C_6}{R^6}+\frac{L_c\left(L_c+1\right)}{2\mu R^2}+E_{\mathrm{thresh}; c}\right]\delta_{c,c'}\\
\label{eq:LRPotential}&-\frac{C_3\left(c,c'\right)}{R^3}\, .
\end{align}
Here, $C_6$ is the van der Waals coefficient, the calculation of which is described in Sec.~\ref{sec:molec-scatt-prop}, $C_3\left(c,c'\right)$ is the coefficient of the anisotropic dipole-dipole interaction~\cite{PhysRevA.84.062703,PhysRevA.81.022702,1367-2630-17-3-035015}, relevant when the two colliding molecules are polar, $L_c$ is the partial wave of the $c^{\mathrm{th}}$ channel, $\mu=m/2$ is the reduced mass of two molecules, $E_{\mathrm{thresh}; c}$ is the threshold energy of channel $c$, which may depend on other parameters, such as an external magnetic field, and the channel index $c$ is understood to encapsulate both internal degrees of freedom such as the rotational angular momentum and rotational projections of the two colliding molecules as well as the external orbital angular momentum and projection.  For illustration, we take these long-range channels to correspond to the rovibrational ground state, so that the number of channels at long range $N_c$ is set by hyperfine degeneracy (black solid lines in Fig.~\ref{fig:MQDT}), and we can ignore the $C_3$ part of the potential in zero electric field.  In contrast, the $C_3$ part of the potential dominates for higher-lying channels which do not correlate to both molecules in their rovibrational ground state~\cite{1367-2630-17-3-035015} (blue dashed lines in Fig.~\ref{fig:MQDT}).  Typical values of the hyperfine degeneracy of a single molecule are $\sim 10-40$ for the alkali dimers~\cite{PhysRevA.88.023605}.  This hyperfine degeneracy is important when considering proper symmetrization of entrance channels, but the energy spacing between nominally degenerate levels is much larger than typical ultracold temperature scales (inset of Fig.~\ref{fig:MQDT}), and so plays little role.  Here, we note that if open channels are allowed to include rotational excitations, for example because molecules have been prepared in a rotational state superposition using an external microwave field~\cite{Ospelkaus_Ni_10,PhysRevLett.116.225306}, the $C_3$ part of the potential should be accounted for in the MQDT.  We leave this modification of MQDT to account for dipolar interactions for future work, and focus on the case in which the open channel manifold has a $1/R^6$ long-range character. 

Expanding the wavefunction in terms of a basis of $N$ states, which includes the hyperfine and partial wave quantum numbers, as
\begin{align}
\psi(R) &=R^{-1}\sum_{c=1}^{N} \Phi_c\left(\Omega\right)\psi_c\left(R\right)\, ,
\end{align}
where $\Omega$ contains all angular and internal degrees of freedom, the radial Schr\"{o}dinger equation becomes
\begin{widetext}
\begin{align}
\sum_{c'=1}^{N}\left[\left(-\frac{1}{2\mu }\frac{d^2}{dR^2}+\frac{L_c\left(L_c+1\right)}{2\mu R^2}+E_{\mathrm{thresh}; c}\right)\delta_{c,c'}+V_{c,c'}\right]\psi_{c'}&=E_c\psi_c\, ,
\end{align}
for a given potential whose matrix elements in the channel space are $V_{c,c'}= \sum_\Omega \Phi_c(\Omega) V(R) \Phi_{c'}(\Omega)$, and $\sum_\Omega$ indicates a ``sum" over all the relevant degrees of freedom.  At long range, any solution of the radial Schr\"{o}dinger equation can be represented by a set of linearly independent pairs of reference wavefunctions $(\hat{f}_c,\hat{g}_c)$, one for each channel, which are solutions of the \emph{uncoupled} long-range potential, i.e.,
\begin{align}
\left(-\frac{1}{2\mu }\frac{d^2}{dR^2}+\frac{L_c\left(L_c+1\right)}{2\mu R^2}+E_{\mathrm{thresh}; c}-\frac{C_6}{R^6}\right)\left\{\begin{array}{c} \hat{f}_c\\ \hat{g}_c\end{array}\right\}&=E_c\left\{\begin{array}{c} \hat{f}_c\\ \hat{g}_c\end{array}\right\}\, ,
\end{align}
\end{widetext}
Since these reference functions are used only as a basis to propagate short-range wave functions out to long-range, we do not have to impose the physical boundary conditions that $\psi_c\to 0$ as $R\to 0$.  Instead, we can choose convenient boundary conditions such that the reference pairs are analytic and only weakly dependent on energy while remaining linearly independent~\cite{ruzic:quantum_2013}.  The energy-analytic reference functions $(\hat{f}_c,\hat{g}_c)$ are connected to the energy-nonanalytic base pair $(f_c,g_c)$ which satisfy physical boundary conditions through the relation 
\begin{align}
\label{eq:MQDTreln}\left(\begin{array}{c} f\\ g\end{array}\right)&=\left(\begin{array}{cc} \mathcal{A}^{1/2}(E)&0\\ \mathcal{A}^{-1/2}(E)\mathcal{G}(E)&\mathcal{A}^{-1/2}(E)\end{array}\right)\left(\begin{array}{c} \hat{f}\\ \hat{g}\end{array}\right)\, ,
\end{align}
where each object is understood to be a matrix/vector in the channel space.  In particular, $\mathcal{A}(E)$ and $\mathcal{G}(E)$ are diagonal matrices in channel space which characterize the MQDT.  In addition, MQDT requires a diagonal matrix of phases $\eta$ specifying phase shifts of $(f_c,g_c)$ relative to free particle solutions as $R\to \infty$.  An optimal MQDT is then constructed using the freedom of the choice of matching radius $\rsr$ such that channels which are asymptotically closed as $R\to\infty$ are classically open at $\rsr$, and so do not contribute energy dependence to the short-range, leading to an energy-analytic reference pair $(\hat{f},\hat{g})$.

The above MQDT prescription captures well the weak energy dependence from channels which are classically open at $\rsr$, and so has been enormously useful for understanding atomic spectra, in which resonances have a relatively low density.  In contrast, the vast number of degrees of freedom in which two molecules can exchange energy at short range leads to a high density of channels that are classically closed at $\rsr$, and impart a rich resonance structure to the short-range physics arising from channels indicated as blue dashed lines in Fig.~\ref{fig:MQDT}.  As discussed in Sec.~\ref{sec:RMT}, the microscopic description of the short-range physics is exceedingly difficult to obtain, and so instead we use a statistical model of resonances based on RMT.  Using the relation Eq.~\eqref{eq:MQDTreln}, the result of propagation of the wavefunction from $\rsr$ through the van der Waals tail of the long-range potential to a large separation $R$ is to modify the coupling strength at energy $\nu_b$ from its ``bare" value $\gamma_{\mathrm{TST}}$  to the value $\gamma=\mathcal{A}(\nu_b)\gamma_{\mathrm{TST}}$~\cite{mayle:statistical_2012}.

For a given partial wave $L_c$, the low-energy behavior of the MQDT parameter $\mathcal{A}(E)$ which modifies the resonance widths at short range is known analytically as~\cite{ruzic:quantum_2013}
\begin{align}
\mathcal{A}(E)&=\left(\frac{\pi 2^{-(2L_c+3/2)}}{\Gamma(\frac{L_c}{2}+\frac{5}{4})\Gamma(L_c+\frac{1}{2})}\right)^2 (\rvdw \sqrt{2\mu E})^{2L_c+1}\, .
\end{align}
Specializing to the $L_c=0$ $s$-wave channel and using the multiplication formula for the Gamma functions, we find
\begin{align}
\label{eq:A-factor-final}\mathcal{A}(E)&=\frac{\Gamma(3/4)^2}{\pi} \rvdw \sqrt{2\mu E}\, .
\end{align}

Of all of the approximations discussed in this work leading to our effective multi-channel Hubbard models, MQDT is likely the most accurate.  This is supported by the exquisite accuracy of MQDT for atomic scattering calculations, and also by its success in atom-molecule scattering problems where a higher density of short-range resonances exist but high-precision, unbiased calculations are still possible~\cite{PhysRevA.84.042703,PhysRevA.86.022711,PhysRevA.90.032711}.

\subsection{Transition state theory to determine bound state-open channel couplings \label{sec:TST}}

In this section we determine the $\gamma_{\text{TST}}$ factor in the dissociation rate $\gamma_b$ (which sets our couplings $\sigma_w$) for two NRMs in a bound state to dissociate to a separation $r=r_{\text{sr}}$. We calculate this rate using TST, a standard approximation used throughout chemistry~\cite{levine:molecular_2010,miller1993beyond,miller2013dynamics}. While widely used, this approximation is probably the least rigorously justified of any in our approach. At the same time, given its broad importance in chemistry, it is likely the most exciting of our approximations to test and explore with ultracold NRMs. Implications if measurements were to confirm or contradict the TST approximation are discussed in Secs.~\ref{sec:beyond-TST} and~\ref{sec:implications-TST-deviate}, as are alternative approximations that could be used in place of the TST.

The
TST dissociation rate $\gamma_b$ is
\be
\gamma_{\text{TST}} &=& \frac{2}{\pi \rho_b}. \label{eq:TST-decay-rate}
\ee
 In particular, this equation is that from the RRKM theory applied to this barrierless reaction (the reaction is unimolecular dissociation\footnote{This nomenclature comes from thinking about the reactant as the BCC and the products the two dissociated NRMs.})
We make the rather strong assumption that TST is valid for all of the bound states, even with large angular momentum. While TST is well-established, at least qualitatively, for averages over many  initial states.
 It is an important chemical question -- which must be answered in order to better understand and control chemical reactions -- to what extent the TST is true for all bound states.
 
Now we have sufficient information to determine the distribution of the couplings $\sigma_w$ from Eq.~\eqref{eq:sigmaw-from-gammab}. Specifically, we use $\gamma_{\text{TST}}$ from Eq.~\eqref{eq:TST-decay-rate} and ${\mc A}$ from Eq.~\eqref{eq:A-factor-final} to determine the decay rate $\gamma_b$ via Eq.~\eqref{eq:decay-from-TST+QDT}. From this, Eq.~\eqref{eq:sigmaw-from-gammab} yields
\be
\sigma_w &=& 2\sqrt{\rvdw/\pi^3\mu \rho_b}\Gamma(3/4).  \label{eq:sigmaw-final}
\ee

This approximation sets the scale for the width of the resonances and typical $U_\alpha$s in Fig.~\ref{fig:beyond-RMT-model-props}(a). 
One qualitative consequence of this is to set the location of the crossover between isolated resonances at small $\omega$ to overlapping resonances. 
This is because when $W_{nb}\ll 1/\rho_b$, the harmonic oscillator state $\ket{n}$ almost always couples to a single bound state, while when $W_{nb}\gg 1/\rho_b$, it couples to many bound states.  
The crossover happens when $W_{nb}\rho_b \sim 1$ or equivalently $(\omega \rho_b)^{3/4}\sqrt{\rvdw(\mu/\rho_b)^{1/2}}\sim 1$. Although the crossover occurs at quite small $\omega$ in Fig.~\ref{fig:beyond-RMT-model-props}(a) (where the tight-binding assumptions of Eq.~\eqref{eq:EffectiveModel} are invalid), for other molecules with smaller $\rho_b$ or $\rvdw$, the crossover will occur at larger $\omega$ where Eq.~\eqref{eq:EffectiveModel}  is accurate. There also may be ways to shift the crossover, for example by shielding the molecules from reaching short range. This could be done using dipolar interactions and anisotropic confinement, analogous to experiments with reactive molecules in Refs.~\cite{Ospelkaus_Ni_10b,ni:dipolar_2010,deMiranda2011} or using other proposed ideas to manipulate the intermolecular interaction~\cite{gorshkov:suppression_2008,wang:tuning_2015,quemener:shielding_2016}.

The key approximation made in TST is that there is a surface in the configuration space of the NRMs such that once the system crosses that surface it never recrosses~\cite{levine:molecular_2010}. TST in its usual formulation also assumes that the dynamics is classical and that prior to undergoing the dynamics of the reaction the system equilibrates inside the dividing surface.
With these assumptions, the RRKM theory allows one to obtain the reaction rate $\gamma_{\text{TST}}$ from knowledge of only the energy at the potential energy minimum and dividing surface, and  the stable vibrational frequencies at these points.

Figure~\ref{fig:TST-cartoon} gives a simple way to understand the RRKM bound state dissociation rate $\gamma_b$ for the present case of barrierless reactions, given by Eq.~\eqref{eq:TST-decay-rate}. The NRMs oscillate in a closed channel -- as a simple picture, imagine this is a harmonic oscillator with frequency $\omega_{\text{cc}}$. Then  this frequency is determined by the density of bound states $\rho_b$; To see this, note that  for a harmonic oscillator the density of states is $\rho_b = 1/\omega_{cc}$, and the oscillations should occur at $\omega_{cc}=1/\rho_{cc}$. Since the density of states for a harmonic oscillator is $\rho_{cc} = 1/\omega_{cc}$, the oscillations should occur at frequency $\omega_{cc}=1/\rho_{cc}$.  With each oscillation, there is an $O(1)$ probability of escaping past the dividing surface, so the dissociation rate is expected to be $\gamma_b \sim \omega_{cc} \sim 1/\rho_b$, as confirmed by Eq.~\eqref{eq:TST-decay-rate}. 

Although the predictions of TST are frequently in accord with experiments studying chemical reactions, there are notable exceptions. Furthermore, even when there is agreement it is murky to what extent the underlying assumptions leading to the TST are valid and to what the TST is valid on a state-by-state basis rather than averaged over many states. The assumption that the dynamics is classical should be accurate for $r<r_{\text{sr}}$ as long as $r_{\text{sr}}$ is chosen small enough. This will ensure that the potential energy is deep in this regime where the TST is being applied, the kinetic energy consequently large, and therefore the dynamics effectively classical (de Broglie wavelength short). The validity of the assumption that inside the dividing surface the system is locally equilibrated is less clear.  

As a consequence of these fundamental questions, it is difficult to assess the likelihood of the TST failing. A priori, large corrections to the TST dissociation rate seem possible or even likely, although perhaps one should hesitate in accepting this conclusion given the accuracy of TST in predicting state-averaged reaction rates. 

In the event that TST requires corrections, the $\gamma_{\text{TST}}$ factor appearing in the expression for $\sigma_w$ will be altered from $2/(\pi \rho_b)$. This will, at the least, change the magnitude of the couplings as determined by $\sigma_w$. In fact, one can imagine that the rate now becomes strongly dependent on the bound state and one must take care to treat different classes of states with different dissociation rates and thus different couplings. For example, large angular momentum states might dissociate more slowly than predicted by TST.

\subsection{Calculating molecular scattering properties: $\rvdw$ and $\rho_b$ \label{sec:molec-scatt-prop}}

As we have seen, within our framework the properties of the Hamiltonian describing the statistical properties of resonant collisions at short range depends universally on the density of resonant states at zero energy $\rho_b$ and the van der Waals length $\rvdw$.  In this section, we provide details on how these quantities are estimated.

\subsubsection{Estimation of $\rvdw$ from the dispersion potential}
We begin with a discussion of $\rvdw$.  $\rvdw$ depends solely on $C_6$, the coefficient of $1/R^6$ in the long-range tail of the potential (see Eq.~\eqref{eq:LRPotential}).  Given two molecules in their electronic ground state and rovibrational states $|v_1N_1M_1;v_2N_2M_2\rangle$, with $M$ the projection of total angular momentum $J$ on a space-fixed coordinate axis, the matrix elements of the $1/R^6$ dispersion potential in the manifold of fixed $\{v_1,N_1,v_2,N_2\}$ arises from second-order degenerate perturbation theory in the dipole-dipole potential $V_{\mathrm{dd}}$ as
\begin{widetext}
\begin{align}
\langle v_1N_1M_1;v_2N_2 M_2|V_{\mathrm{disp}}\left(\mathbf{R}\right)|v_1N_1 M_1'; v_2N_2M_2'\rangle&=-\sum_{\gamma_1,\gamma_2}'\frac{\langle v_1N_1M_1;v_2N_2M_2|V_{\mathrm{dd}}|\gamma_1;\gamma_2\rangle\langle \gamma_1;\gamma_2|V_{\mathrm{dd}}|v_1N_1M_1';v_2N_2M_2'\rangle}{E_{\gamma_1}+E_{\gamma_2}-E_{v_1N_1M_1}-E_{v_2N_2M_2}}\, ,
\end{align}
\end{widetext}
where prime on the summation over $\gamma_1$ and $\gamma_2$ indicates all states, including continuum states and electronic excitations, whose combined energies $(E_{\gamma_1}+E_{\gamma_2})$ are non-degenerate with the energies of the degenerate manifold $(E_{v_1N_1M_1}+E_{v_2N_2M_2})$.  The dipole-dipole potential has the well-known form
\begin{align}
V_{\mathrm{dd}}&=\frac{\mathbf{d}_1\cdot\mathbf{d}_2-3\left(\mathbf{d}_1\cdot\mathbf{e}_R\right)\left(\mathbf{d}_2\cdot\mathbf{e}_R\right)}{R^3}\, ,
\end{align}
with $\mathbf{d}_i$ the dipole operator of molecule $i$, $R=|\mathbf{R}|$ the intermolecular separation, and $\mathbf{e}_R=\v{R}/R$ a unit vector in the direction of $\mathbf{R}$.  Because of the ``bilinear" form of $V_{\mathrm{dd}}$ on the dipole operators, the expectations of $V_{\mathrm{dd}}$ can be related to expectations of diagonal elements of the dynamical polarizability tensor evaluated at pure imaginary frequency $\tilde{\alpha}(i\omega)$, defined as
\begin{align}
\nonumber\langle vN M|\alpha_{qq}\left(i\omega\right)|vNM\rangle&=\sum_{\gamma}'\frac{E_{\gamma}-E_{vNM}}{\left(E_{\gamma}-E_{vNM}\right)^2+\left(\hbar\omega\right)^2}\\
 &\times \left|\langle \gamma|\mathbf{d}\cdot\mathbf{e}_q|vNM\rangle\right|^2\, .
\end{align}
To wit, the isotropic scalar part of $C_6$ may be written as
\begin{align}
\label{eq:IsoC6}C_6^{\mathrm{iso}}&=\frac{3\hbar }{\pi} \int_0^{\infty} d\omega \prod_{j=1}^2 \langle v_j N_j M_j|\bar{\alpha}(i\omega)|v_jN_jM_j\rangle \, ,
\end{align}
where $\bar{\alpha}(i\omega)=\mathrm{Tr}(\tilde{\alpha}(i\omega))/3$.  This isotropic $C_6$ is the only contribution to the dispersion potential for the rotational ground state, but for rotationally excited states there is an additional isotropic contribution whose operator character is that of the scalar product of two rank-two operators, as well as anisotropic contributions with rank-two operator character.  The coefficients of these additional contributions can be written in a form similar to Eq.~\eqref{eq:IsoC6} but additionally involving the polarizability tensor invariant $\Delta \alpha=\sqrt{(3\mathrm{Tr}(\tilde{\alpha}^2)-(\mathrm{Tr}\tilde{\alpha})^2)/2}$, which together with $\bar{\alpha}$ completely specifies the polarizability tensor for diatomic molecules.  Since in this paper we focus only on the case in which the open channel consists of hyperfine states within the rovibrational ground state manifold, we do not explicitly give expressions for the anisotropic parts of the dispersion potential here, but refer the reader to Ref.~\cite{1367-2630-12-7-073041}.

To calculate the expression Eq.~\eqref{eq:IsoC6} for the isotropic $C_6$ coefficient, one needs the energies and transition dipole moments of the molecule in question.  These quantities are usually obtained via electronic structure calculations, for example multi-reference configuration interaction methods~\cite{kotochigova2005ab,1.2817592}.  For molecules with $^1\Sigma$ ground states, which includes all of the alkali dimers, the relevant contributions to the polarizability come from transitions to the lowest-lying $^1\Sigma$ and $^1\Pi$ electronic states.  Since these levels do not require relativistic spin-orbit coupling in order to obtain a dipole moment (in contrast to, say, the $^3\Pi$ level), non-relativistic electronic structure calculations can be employed.  As an order of magnitude estimate, one can approximate the electronic and rotational contributions to $C_6$ by the Uns\"{o}ld approximation $3 U\bar{\alpha}^2/4$~\cite{stone2013theory} and $d^4/6B$, respectively, where $U$ is a mean excitation energy, $d$ the permanent dipole moment, and $B$ the rotational constant.  For typical alkali metal dimers, taking $U\sim 1$eV$\approx0.05$ atomic units~\cite{julienne:universal_2011}, this gives an order of magnitude of $C_6\sim 10^4-10^5$ atomic units, where 1 au=$E_ha_0^6$, with $E_h$ the Hartree energy and $a_0$ the Bohr radius.  Putting in realistic numbers, this crude estimate is off from the \emph{ab initio} calculations by 10-20\%.  Taking this as a crude estimate for the precision with which we know $C_6$, we can estimate that our calculation of $\mathcal{A}(E)\propto C_6^{1/4}$, the MQDT parameter responsible for narrowing of short-range resonance due to threshold scattering effects, is off by five percent or less.

\subsubsection{Estimation of the density of rovibrational states at zero energy}
We now turn to as estimation of $\rho_b$, the density of states (DOS) at zero energy. We outline the calculation performed in Ref.~\cite{mayle:statistical_2012,mayle:scattering_2013}, and then we show that one can obtain the order of magnitude of $\rho_b$ by a simple analytic formula.  The calculation in Refs.~\cite{mayle:statistical_2012,mayle:scattering_2013}   used the model potential
\begin{align}
\label{eq:EstPot} V(R)&=V_{\mathrm{LJ}}\left(R\right)+\frac{L_c\left(L_c+1\right)}{2\mu R^2}+E_{v_1,N_1,v_2,N_2}\, ,
\end{align}
where $V_{\mathrm{LJ}}\left(R\right)=-C_6/R^6 + C_{12}/R^{12}$ is a Lennard-Jones (LJ) potential with the $C_6$ calculated as above and $E_{v_1,N_1,v_2,N_2}$ is the threshold energy of the channel with molecules in vibrational states $v_1,v_2$ and rotational states $N_1,N_2$.  The depth of the potential Eq.~\eqref{eq:EstPot} was set to give the correct binding energy of the bimolecular complex relative to the free-molecule threshold.  One then calculates the bound states for each channel independently, and obtains the total density of states by counting the bound state from all of the channels.
A key assumption employed by Refs.~\cite{mayle:statistical_2012,mayle:scattering_2013} when constructing the DOS is that each state that preserves both the energy and angular momentum is counted.  This, in some sense, is equivalent to the argument that the classical phase space during a collision at short range is ergodically sampled, which has been verified in classical trajectory simulations of diatom-diatom collisions~\cite{PhysRevA.89.012714}.  

One can go beyond this fully ergodic assumption by while retaining the same general picture by specifying which set of states should couple. Then one simply obtains a reduced density of states. One approach to restricting the states is to consider a cutoff in the maximum angular momentum $J$ that is allowed.   Even when the condition that every state contributes is relaxed, the density of resonances near zero energy for a colliding pair of molecules is many orders of magnitude larger than for atoms, approaching several thousands per Gauss of magnetic field (where $1G\sim 100$nK$\sim 2\times 2\pi$kHz in energy units) for heavy molecules like RbCs~\cite{mayle:scattering_2013}.  This should be contrasted with typical alkali atoms, including mixtures, for which $\rho\sim 1/100 G^{-1}$~\cite{PhysRevA.87.032517,PhysRevA.85.032506}.

The means of estimating $\rho_b$ above should not be taken as anything more than an order of magnitude estimate.  There are several reasons for this.  First, the potential Eq.~\eqref{eq:EstPot} employed in the estimate, while making some connection with well-estimated quantities like $C_6$ and the binding energy, remains a model potential and will not have the same level density as the true potential.  Additionally, while many of the states at zero energy will participate during a collisional event, not all of them necessarily do for each event over the relevant timescales.  

In addition to the above detailed treatment, it is useful to have a simple scaling argument for the density of states.  For a heteronuclear diatomic molecule, the density of rotational states for a single molecule is $\rho=1/B$.  Using this uniform density of states, we then have that the number of ways for two molecules to have total energy $E$ is $E/B$.  To see this, consider that the uniform density of states corresponds to an equally spaced spectrum (harmonic oscillator) indexed by quantum numbers $n_1$ and $n_2$, and then note that $E=B(n_1+n_2)$.  Now, each of the channel energies is lowered by the potential depth $D$ when two NRMs reach short range, and so the relevant excitation energy for zero collision energy is $E=D$.  

So far, we have only included the contributions from rotational states, and so we should multiply by the number of bound states per two-molecule channel $M$ and the number of vibrational excitations per molecule $X$.  One can expect that the number of vibrational excitations scales as $\sim 1/D$, and indeed using $D_{\mathrm{RbCs}}=800$ cm$^{-1}$ and $D_{\mathrm{KRb}}=2779.6$ cm$^{-1}$~\cite{mayle:scattering_2013} together with $X_{\mathrm{RbCs}}=129$~\cite{PhysRevA.83.052519} and $X_{\mathrm{KRb}}\sim 30$~\cite{PhysRevA.76.022511}, we find $D_{\mathrm{RbCs}}X_{\mathrm{RbCs}}\approx D_{\mathrm{KRb}}X_{\mathrm{KRb}}$ within 20\%.  We next estimate the dependence of the number of bound states per two-molecule channel, $M$, on $D$. The depth of the two-molecule potential is roughly $V_{\mathrm{min}}=6 D$, as there are $\binom{4}{2}=6$ pairwise connections between the atoms at short range.  We can approximate the two-molecule potential as a LJ potential with this depth, 
\begin{align}
\tilde{V}_{\mathrm{LJ}}\left(R\right)&=\frac{1}{2\mu R_{\mathrm{vdW}}^2}\left[\lambda^6\left(\frac{R_{\mathrm{vdW}}}{R}\right)^{12}-\left(\frac{R_{\mathrm{vdW}}}{R}\right)^{6}\right]\, ,
\end{align}
where $\lambda=1/(\sqrt{2}R_{\mathrm{vdW}}^{1/3}V_{\mathrm{min}}^{1/6}\mu^{1/6})$.  If we now use Levinson's theorem, $M=[\phi(E=\infty)-\phi(0)]/\pi-1/2$, to estimate the number of bound states. To obtain the phase difference, we use a WKB approximation for the phase. In particular, we calculate the phase associated with propagating from the left-most classical turning point $R=\lambda \rvdw$ to infinity. (We cannot calculate the phase from $R=0$ to $\infty$ because this diverges for the LJ potential. This is an artifact of LJ, and instead we simply realize that the phase accumulated up to the left-most turning point is small in the real potential.) Using this phase, we find the number of bound states is 
\begin{align}
M&\approx \frac{1}{\lambda^2 \pi}\int_{1}^{\infty}dz \sqrt{\left[\left(\frac{1}{z}\right)^{6}-\left(\frac{1}{z}\right)^{12}\right]}\, ,\\
&=\frac{\Gamma(\frac{1}{3})}{6^{2/3}\sqrt{\pi} \Gamma(\frac{11}{6})}D^{1/3}R_{\mathrm{vdW}}^{2/3}\mu^{1/3}\, .
\end{align}
Putting this all together, we obtain the estimate
\begin{align}
\rho_b&\sim \frac{M X D}{B^2}\, ,
\end{align}
which scales roughly as
\begin{align}
\label{eq:rhobscal}\rho_b&\sim \frac{D^{1/3}R_{\mathrm{vdW}}^{2/3}\mu^{1/3}}{B^2}\, ,
\end{align}
where the scaling relation $\sim$ includes a factor with units of energy that does not scale with any of the factors on the right hand side.

\section{Beyond the standard suite of approximations \label{sec:beyond-standard-approxes}}

Section~\ref{sec:approx} applied our ``standard suite of approximations" to determine the effective model parameters ${\mc O}_\alpha$ and $U_{\alpha}$.
The present section evaluates this  suite of approximations' accuracy and go beyond it. We discuss the likelihood that the assumptions behind the approximations are satisfied for NRMs. We also delineate consequences for the effective lattice model if an approximation fails, for example if the modification is expected to merely be a small quantitative shift or if it introduces wholly new features. We introduce methods to incorporate physics  beyond each approximation, with models  motivated by a combination of experimental and theoretical knowledge of these systems, as well as mathematical simplicity.   

It will be exciting to compare theory to ongoing experiments, regardless of outcome, as we discuss for each approximation. On the one hand, quantitative agreement would confirm the derived lattice model and the approximations behind it, and this would provide a solid basis for future experiments controlling chemistry and exploring many-body physics. On the other hand, a discrepancy would teach us something surprising about  strongly-held assumptions regarding intermolecular interactions, quantum chaos, or chemical kinetics, indicating the need for new perspectives. For example, TST is widely believed to adequately describe reaction rates in a wide variety of molecules~\cite{levine:molecular_2010}, so a discrepancy would have fundamentally important consequences in chemistry. Similarly, RMT is believed to govern chaotic scattering~\cite{PhysRevLett.52.1}, so a discrepancy would indicate that scattering molecules behave either more regularly than anticipated or that the link between classical chaos and a RMT description of the quantum system is more restricted than expected.

\subsection{Separation of length scales: beyond extreme separation \label{sec:length-sep-beyond}}

\subsubsection{What happens if separation of length scales is not extreme}

For typical molecules, $\rsr<\rvdw<\lho$, and each inequality holds by a factor of $\sim5$. For example, for the numbers quoted in the introduction to this section these inequalities are $4\text{nm}<25\text{nm}<100\text{nm}$. Therefore, the conclusions derived resulting from the assumption that $\rsr \ll \rvdw \ll \lho$  are expected to be qualitatively valid. 
For example, the structure of  many resonances with repelled levels and a smooth distribution of couplings with  finite variance, should still survive.
However, quantitative effects might be very naively estimated at the $\sim 1/5 =20\%$ level.  These quantitative effects might include, for example, some weak dependence on harmonic oscillator quantum number $n$. 
These quantitative modifications can arise from either $r_{\text{sr}}/\rvdw$  or $\rvdw/\lho$ being non-negligible.
 
\textbf{Effects of $r_{\text{sr}} \centernot\ll \rvdw$.}
Three assumptions may need to be modified in this case. The first two modifications occur if $R$ in Region 1 becomes comparable to $\rvdw$, which happens if $r_{12}$ is too large. The third modification occurs if $R$ in Region 2 becomes comparable to $r_{\text{sr}}$, which happens if $r_{12}$ is chosen to be too small. 
The first assumption that can fail is that the  
BCC's dissociation dynamics is classical and therefore can be treated with TST. 
This assumption can fail if  $R$ in Region 1 becomes comparable to $\rvdw$,  since in this condition the vdW potential may be sufficiently shallow that the small kinetic energy leads to a de Broglie wavelength that is not negligible compared to the length scale on which the potential varies. 
The second assumption that can fail
is that the dynamics is chaotic in Region 1 and RMT applies. If $R$ becomes too large, this assumption fails.    
The third assumption that can fail is that the potential in Region 2 is purely vdW and can be treated with QDT. If $R$ in this region are allowed to be too small, the vdW potential is no longer an accurate approximation. 

Individually, these effects can be minimized by proper choice of $r_{12}$, but there is a tradeoff:  larger $r_{12}$ ensure that the potential in Region 2 is purely vdW, but can make the dynamics inside $r_{\text{sr}}$ less chaotic and classical and RMT and TST less applicable; smaller $r_{12}$ ensure that the dynamics in Region 1 is chaotic and classical, but can make the potential in Region 2 more complicated than the vdW potential.

The consequences of these failures
depend on the reason for the failure. If the failure is due to the invalidity of TST to treat the out-flow from Region 1 to Region 2 due to quantum effects, then the TST must be augmented to include these. If the dynamics is not chaotic, the RMT must be modified.  If the failure is due to the invalidity of treating the potential in Region 2 as a vdW potential, then one must modify the QDT to account for the more complicated correct potential. 
The breakdowns of TST and QDT can modify $\sigma_w$ and its dependence on the energy of the bound states.    
This in turn will change the $U_\alpha$s and ${\mc O}_\alpha$s appearing in the lattice model by changing the range of $\omega$ over which the $U_\alpha$ varies, i.e. the ``width" of the resonance-like features in Fig.~\ref{fig:beyond-RMT-model-props}(a). 
The breakdown of RMT can lead to altered distributions of bound state energies, as well as modified coupling magnitude $\sigma_w$. 
The breakdown of RMT, TST, and QDT in these ways -- and methods to do calculations when they break down -- are discussed in more detail in Secs.~\ref{sec:RMT}, \ref{sec:TST} and~\ref{sec:QDT}, respectively.

\textbf{Effects of $\rvdw \centernot\ll \lho $.}
Two assumptions may break down in this case. Since $\rvdw \centernot\ll \lho $, $r_{23}$ must either fail to satisfy $r_{23}\gg \rvdw$ or $r_{23}\ll \lho$. However, at most one of these need fail: For example, even if $r_{23}\centernot \gg \rvdw$, we can still choose $r_{23}\ll \lho$. 

If $r_{23}$ is too large, so that $r_{23}\centernot \ll \lho$, the propagation in Region 2 can no longer incorporate only the vdW potential, but must also include the harmonic oscillator. This additional potential will modify the QDT that is used to calculate the wavefunctions in this region. This will lead to an  energy dependent correction to $\sigma_w$, resulting in additional dependence of the $U_\alpha$ and ${\mc O}_\alpha$'s  on principal quantum number and trap frequency $\omega$.  These effects may be included in our calculations by numerically solving for the QDT parameters by incorporating the full vdW plus  harmonic oscillator potential, as described in Sec.~\ref{sec:QDT}. Because this numerical solution is for  a single channel potential in the radial coordinate, it can be carried out in reasonable computer time using standard algorithms.

On the other hand, if $r_{23}$ is too small, so that $r_{23}\centernot \gg \rvdw$, the eigenstate in Region 3 can no longer be solved as the harmonic oscillator coupled to zero-range states, but must also include the vdW potential.  
 As before, the physical consequence will be that for two NRMs in a trap, the eigenstates and energies gain a modified dependence on principal quantum number and oscillator frequency $\omega$. Resultantly, so do the $U_\alpha$s and ${\mc O}_\alpha$s. These effects may be incorporated in our calculations by replacing the harmonic oscillator eigenstates with the numerical solutions of the harmonic plus vdW potential in the presence of the appropriate short range couplings (which are still given by TST+QDT applied to Regions 1 and 2). Again, because this calculation is for a single channel potential in the radial coordinate, it can be carried out efficiently.

\subsubsection{Conclusions: separation of lengths}

If the separation $\rsr \ll \rvdw \ll \lho$ is insufficiently extreme, then our eigenstates and eigenergies may be modified with additional principal quantum number and $\omega$-dependences, and this will be reflected in the $U_\alpha$ and ${\mc O}_\alpha$ in the effective lattice model Eq.~\eqref{eq:EffectiveModel}. The effects will depend on the precise nature of the overlap of length scales, but in all cases, the effects should be moderate ($\lsim 20\%$)  and can be included numerically exactly by modifying the Region 2 (QDT) or Region 3 solutions to incorporate the appropriate effects. In all cases, one must numerically solve a modified single channel potential, a straightforwardly tractable problem.

We emphasize that whatever corrections are present for NRMs in a lattice, they are not expected to be substantially larger or different in character than those occurring for ultracold \textit{atoms}. For atoms, again, one needs the short range and vdW lengths to be much smaller then $\lho$ in order for the short-ranged pseudopotential to be a valid approximation ot the true potential, and consequently for the standard Hubbard model and expressions for $U$ to be valid~\cite{jaksch_bruder_98}. Corrections to these approximations have been predicted for atoms~\cite{PhysRevA.90.043631,PhysRevLett.105.170403}, and often are small but may be quantitatively important. 

The effects of the overlapping length scales -- for both NRMs and atoms -- are likely to be exaggerated in experiments in deep microtraps~\cite{schlosser2001sub,weiss2004another,miroshnychenko2006quantum,beugnon2007two,PhysRevA.88.013420,1367-2630-13-4-043007,grunzweig2010near,serwane2011deterministic,Kaufman2012,PhysRevX.4.021034,Kaufman_Lester_14}. Such experiments are an exciting route for creating and exploring ultracold molecules~\cite{hutzler2016eliminating}, and in them the  $\lho$ can in principle be greatly reduced by taking of advantage of the large trap depths that are available.

\subsection{Beyond random matrix theory for BCC energies  \label{sec:RMT-beyond}}
 
\subsubsection{Beyond RMT} 

Although it is compelling that RMT should govern the distributions in Eqs.~\eqref{eq:sr-ham-prob-eq} and~\eqref{eq:wb-gaussian}, the applicability of RMT rests on a couple of assumptions. The principle assumption is  that the classical short-range dynamics is  chaotic. There is some evidence for this~\cite{PhysRevA.89.012714,kovacs1995chaotic,PhysRevE.79.026215,atkins1995classical,Pattard1998360,PhysRevLett.89.203202,PhysRevE.78.046204}.
Ref.~\cite{PhysRevA.89.012714}, found chaos in the short-range dynamics of atom-diatom collisions. The collisions between NRMs would naively be expected to be  more chaotic, although this has not yet been confirmed. More importantly, the dynamics is seen to be chaotic only within some range of intermolecular separation. If we choose $r_{\text{sr}}$ as the boundary within which the dynamics is chaotic and it is valid to apply the RMT (see Sec.~\ref{sec:length-sep}), and $r_{\text{sr}}$ is too small, then  the regions $R>r_{\text{sr}}$ will involve physics that we have neglected in our other approximations. For example, if $r_{\text{sr}}$ is too small, the potential for $R>r_{\text{sr}}$ will involve contributions more complex than the simple vdW potential. Furthermore, although the general connection between classically chaotic dynamics and RMT is well-established, precise quantitative connections between these in complex quantum systems are few, so there is potential for surprising phenomena.  

If RMT fails, it is likely that the qualitative structure persists -- many bound states, with a $\rho_b$ similar to the current estimates, may influence the long-range physics -- but that the distribution of levels would be altered. For example, the spacing distribution might not reproduce Fig.~\ref{fig:nearest-level-prob-dist}(a). Similarly the couplings $w_b$ might fluctuate with roughly the scale given by Eq.~\eqref{eq:wb-gaussian}, but with a distribution that quantitatively deviates from a Gaussian.  It is not unthinkable that in special cases these corrections could be significant, especially for special lattice depths (controlling the collision energy), in an ``accidentally regular" molecule, or at certain values of external (electric or magnetic) fields. 

In the remainder of this section we propose and examine the consequences of a few models of the level distributions and couplings  that could describe the physics even when the RMT is invalid. Lacking quantitative microscopic models for the $\ket{b}$s and $\nu_b$s beyond the RMT, we are guided by qualitative considerations, mathematical simplicity to capture basic features, and analogies to other systems with complex scattering where more is known. The primary analog systems are provided by complex ultracold atoms, such as the lanthanides Dy~\cite{PhysRevLett.104.063001,PhysRevLett.107.190401,1367-2630-17-4-045006} and Er~\cite{PhysRevA.85.051401,PhysRevLett.108.210401}, where recently the complex collisional physics has been explored~\cite{PhysRevA.89.020701,frisch2014quantum,maier2015broad,maier2015emergence}. These atoms have a much denser set of bound states than alkali atoms, leading to coupling of many scattering channels during the collision and thus RMT-like distributions of levels. However, they have a much lower density of bound states than NRMs, such that typically only a single level should be coupled to at ultracold temperatures or in lattices of typical depth.

The Er and Dy systems that we take as analogs show clear signatures of complex scattering: the level spacing distribution is not that of independently distributed levels [compare Fig.~\ref{fig:nearest-level-prob-dist}(b)]. Rather the levels are repelled analogous to the RMT. However, some discrepancies from the RMT's GOE distribution are present. First, although there is clear level repulsion relative to that found for independent levels, it is less strong than in the GOE. The spacing distribution is relatively well fit by a Brody distribution that interpolates between the Poisson and GOE limits\footnote{The distribution has an interesting magnetic field dependence~\cite{PhysRevA.89.020701}.}~\cite{frisch2014quantum,maier2015emergence}.  Another discrepancy is that, in some ranges of external magnetic field, there are a large number of bound states with an RMT-like distribution coexisting with a single, orders-of-magnitude broader level~\cite{maier2015broad}.

Inspired by these analogies with lanthanides and to capture other simple deviations from RMT, we calculate the effective Hubbard model parameters ${\mc O}_\alpha$ and $U_\alpha$, modeling bound state properties beyond RMT using three non-RMT models. We consider a Poisson distribution, an RMT coupled to a single broad level, and two interleaved RMT ensembles with different $\rho_b$.

\textbf{Poissonian statistics}. The first alternative model to the GOE is to sample $N_b$ eigenvalues, 
taking each $\nu_b$ from a uniform distribution 
\be
P_{\text{Pois}}(\nu_b) &=& \begin{cases} \frac{\rho_b}{N_b} &\hspace{0.35in} \text{if } -\frac{N_b}{2\rho_b} < \nu_b < \frac{N_b}{2\rho_b}   \\
0 &\hspace{0.35in} \text{otherwise}
\end{cases}\label{eq:poisson-nu-b}
\ee
for  $N_b\rightarrow \infty$ while fixing $\rho_b$. 
This distribution gives $N_b$ levels randomly each sampled from a uniform distribution of width $N_b/\rho_b$, and thus gives an average density of eigenvalues $\rho_b$.  
We refer to this as a Poisson distribution, following convention\footnote{This (fairly standard) nomenclature is used because it is a distribution of spacings arising from a Poisson process. It is not to be confused with the usual Poisson distribution, which counts a discrete number of events $k$ in a certain time interval, where the events are sampled from a Poisson process. (These two distributions do both take the form of exponentials of their argument, however.)}. Such a distribution may be expected to describe non-chaotic (integrable) systems, where there are a large number of conserved quantities.
The distribution Eq.~\eqref{eq:poisson-nu-b} leads to the exponentially decaying level spacing distribution shown Fig.~\ref{fig:nearest-level-prob-dist}(b). The lack of level repulsion relative to the GOE is apparent.

The distribution of the $w_b$s for the case of Poisson level statistics is less constrained than it was for the GOE. For the GOE, the symmetry under orthogonal transformations restricted the probability distribution of $w_b$ to a Gaussian independent of the other $w_b$s. No such symmetry requirement constrains the Poisson case. The $w_b$s will depend on the microscopic model. We nevertheless expect them to be governed by a distribution of some finite variance. For the purpose of making illustrative plots, we simply use the same distribution of $w_b$s as for the GOE, Eq.~\eqref{eq:wb-gaussian}.

Fig.~\ref{fig:beyond-RMT-model-props}(b) shows the structure of the $E_\alpha$s, $U_\alpha$s, and ${\mc O}_\alpha$s for the Poisson distribution, alongside the results reproduced  for the usual GOE in Fig.~\ref{fig:beyond-RMT-model-props}(a). The results are very similar, though with careful analysis, one can see that the bound states and associated resonances in the Poisson ensemble are not repelled as in the GOE, and consequently have a less regular spacing.
 
\textbf{Broad resonance in RMT background.} A remarkable feature of complex atomic scattering has emerged from studies of lanthanide atom collisions~\cite{maier2015broad}: almost all resonances (bound states) are part of a dense forest of narrow, roughly RMT-distributed levels, but occasionally an orders-of-magnitude broader resonance appears and couples to this background. We mimic this by adding a single  additional level at energy $\nu'$ with coupling strength $w'$. 

A strong hypothesis has been advanced in Ref.~\cite{maier2015broad} for the physical origin of this effect. They suggest that the closed interaction channels couple to create the dense set of RMT-like resonances, while the open channel harbors a shallow bound state that does not couple to  the closed channels strongly enough to inherit the RMT properties. This shallow bound state has universal properties (e.g., dependence on $B$-field) that are standard for single-channel scattering. This hypothesis is able to account for the data, but its microscopic origin -- why the open channel bound state is not described with RMT-like statistics -- remains unclear. 

One way of thinking about this hypothesis semiclassically is that part of the phase space has chaotic dynamics (giving rise to the dense RMT bound states), while some small portion of the phase space has integrable dynamics (giving rise to the singled out bound state). This is reminiscent of the phenomenon of ``quantum scarring" in quantum systems where the phase space distribution is invariant over a large region of phase space, as expected for chaotic, ergodic dynamics, but certain ``scarred" regions display more interesting 
structure~\cite{heller:bound-state_1984}.

Fig.~\ref{fig:beyond-RMT-model-props}(c) shows the effects that this broad resonance~+~RMT model has on the $E_\alpha$s, $U_\alpha$s, and ${\mc O}_\alpha$s. It shows that the resonances in $U_\alpha$ are qualitatively like the GOE with a broad resonance in $U_\alpha$s additively superposed with this. 

\textbf{Two overlapping RMT distributions.} One may extrapolate into a new regime the idea that there are separate regions in phase space that are uncoupled: Rather than an integrable piece and a chaotic piece as considered above, we can consider two chaotic pieces which are weakly coupled. This would occur for example, if there were a nearly-conserved quantum number, but chaotic dynamics within each manifold of that conserved quantum number. One could imagine a scenario where the underlying microscopic separation is natural in terms of the usual degrees of freedom -- for example one could imagine that rotational sectors mix strongly, but vibrational  sectors couple only weakly. Alternatively,  the regions of phase space that decouple could be highly nontrivially related to the rotational and vibrational states.

Although there is no obvious such separation in the dynamics of two colliding NRMs, 
it is a plausible route to a failure of the usual RMT, and it is mathematically simple to capture.
If there are uncoupled or weakly coupled ergodic regions, we expect each of the two Hilbert spaces corresponding to these regions to inherit their own RMT. That is, we take a realization of $\nu_b$s to be the 
union of two sets labeled by $j=1,2$, each of which samples eigenvalues of ${\hat H}_b$ and couplings $w_b$ from the  GOE probability distributions
\be
P^{(j)}(\Hb) &=& {\mc N}^{(j)}_H e^{-\operatorname{Tr}{\Hb}^2/2\sigma_j^2} \\
P_w^{(j)}(w_b) &=& {\mc N}^{(j)}_w e^{-w_b^2/2\sigma_{w,j}^2}. \label{eqs:2-RMT}
\ee
The probability distributions for $j=1$ and $j=2$ are distinct
because the values of $\sigma_j$ are distinct (i.e. $\sigma_1\ne \sigma_2$).  Note that the single broad resonance + RMT distribution considered previously emerges as the limit of this 2-RMT distribution when the $\rho_b^{(1)}/\rho_b^{(2)}$ differs greatly from unity. Then, for a given energy window, there will be many levels from one of the distributions that are relevant, while the other distribution will contribute a single level in this window. 

Fig.~\ref{fig:nearest-level-prob-dist}(c) shows the nearest-level spacing distribution for the two-GOE ensemble. The $R_{\text{vdW}}$, $\mu$, and $\rho_b^{(1)}$ are  identical to the standard GOE case, while the second GOE has $\rho_b^{(2)}=0.1\rho_b^{(1)}$.
We emphasize that sampling the $\nu_b$'s as a union of two GOE ensembles with standard deviations
$\sigma_1$ and $\sigma_2$ is not equivalent 
to sampling the $\nu_b$'s from a single GOE ensemble $\sigma_{\text{tot}}$, regardless
of how $\sigma_{\text{tot}}$ is chosen\footnote{To convince yourself that this
is true, consider $\sigma^{(1)}\ll \sigma^{(2)}$: the first distribution's levels are much closer together than the second's. The probability distribution of 
nearest-level spacings $\Delta$ is then dominated for not-too-large $\Delta$ by the
contributions from $P^{(1)}$, since it is unlikely that a level from $P^{(2)}$ will 
be close. However, the large $\Delta$  tail of the distribution must decay faster 
than $P^{(1)}$: the probability of the nearest-level from $P^{(1)}$ being at least 
$\Delta$ is small, but the probability of the nearest-level from \textit{either}
 distribution being at least $\Delta$ is much smaller, since it would require there 
being no level within $\Delta$ from either $P^{(1)}$ or $P^{(2)}$.}.  
Fig.~\ref{fig:nearest-level-prob-dist}(c) illustrates one aspect of how the two-RMT distribution differs from the usual GOE RMT distribution by contrasting the shape of the nearest-level 
probability distribution in each case.  We also note that the neighbor spacing distribution is not the superposition of the spacing distribution for each of the RMT distributions (although in some cases this is true or a good approximation~\cite{mehta2004random}).

Fig.~\ref{fig:beyond-RMT-model-props} shows the $E_{\alpha}$s, ${\mc O}_\alpha$s, and $U_\alpha$s for the two-GOE ensemble. The structure of $U_\alpha$s is a forest of resonances from the dense GOE superposed with broader resonances resulting from the less dense GOE.

\textbf{Other distributions.} The three beyond-RMT models we have introduced are illustrative. There are obvious extensions that take the considerations above further. For example, one could interpolate between the Poisson and GOE limits, perhaps using a Brody distribution~\cite{brody1973statistical}. Or one could interpolate these by deviating from a completely random GOE ensemble through adding more and more integrals of motion~\cite{PhysRevE.93.052114}.  Another direction is that  one could include three, four,  or more overlapping RMTs. A final approach is that, in principle, one can compute the detailed level statistics and couplings from the microscopic four-atom problem using the techniques in Ref.~\cite{wall:microscopic-derivation-multichannel_2016}. Reference~\cite{wall:microscopic-derivation-multichannel_2016} sets this up formally, although the calculations would be challenging already for simple diatomic molecules.  

\subsubsection{Exploring RMT and beyond in experiments.} 
Quantitatively exploring the nature of the levels and their couplings in would provide valuable insight into quantum chaos. Although the connection between classical chaos and RMT spectral statistics is well-established both theoretically and experimentally, never has it been possible to quantitatively explore the phenomena in systems possessing the multitude of tuning knobs that NRMs offer -- e.g., molecular species, external electric fields, magnetic fields, and optical lattice depths.

\subsection{Beyond transition state theory for the molecular dissociation rate $\gamma_b$  \label{sec:TST-beyond}}

\subsubsection{Beyond TST \label{sec:beyond-TST}}

There is no universally applicable approximation to capture corrections to TST, and the theory used will depend on the relevant phenomena that one aims to capture. For example, calculating quantum corrections   will require different techniques than calculating the effects of locally being out of equilibrium. Accounting for recrossing across the dividing surface similarly require a different set of approaches.

One of the most satisfying approaches would be to compute the dissociation rate from a model of the constituent atoms. This would be captured via the approach in Ref.~\cite{wall:microscopic-derivation-multichannel_2016}. However, this is a computationally formidable problem. Quantitatively calculating reaction rates  -- in a way that is reliable without independent confirmation -- is possible only for the simplest molecules. 

Absent such a microscopic approach, it is necessary to employ less controlled approximations. A phenomenological approach is to consider the true dissociation rate to be some numerical prefactor times $\gamma_{\text{TST}}$. For example, if one expects significant recrossings of the dividing surface, one may expect a reduced dissociation rate. On the other hand, if one is concerned that the system may be out of equilibrium within the dividing surface, then the full phase space may not be explored and the effective $\rho_b$ is reduced. It is straightforward to phenomenologically account for these features, but quantitatively calculating these factors is challenging.

One potentially crucial effect in reaction dynamics is  that intramolecular vibrational energy redistribution (IVR) proceeds at a finite rate, and may take longer than the reaction itself. This idea has succeeding in explaining several non-RRKM unimolecular dissociation reaction rates. 
One prominent approach to including the finite rate of IVR over the reaction timescale is the local random matrix theory (LRMT)~\cite{logan1990quantum,schofield1995rate,leitner1997vibrational,manikandan2014dynamical,gruebele2004vibrational,lourderaj2009theoretical,farantos2009energy}.  This accounts for the fact that locally in space the system may be ergodic and described by RMT, but that energy transfer spatially can be slow compared to the timescale of the reaction. One can observe enhancements or reductions of the RRKM dissociation rate by an order of magnitude. It is an interesting question to what extent this finite time for energy transport will be relevant for small, diatomic molecules versus large, complex ones.

\subsubsection{Implications if experiments measure deviations from TST \label{sec:implications-TST-deviate}}
 
The TST is the most uncontrolled approximation used in our calculations, and consequently testing this approximation and clarifying its applicability will potentially provide a huge scientific payoff. This is because the TST is a crucial tool in chemical kinetics, from atmospheric chemistry to pharmaceuticals~\cite{levine:molecular_2010}. Numerous comparisons of TST to experiment exist, but rarely with the versatility and control offered by ultracold NRMs. Additionally, due to the temperatures involved, prior experiments invariably considered the reaction rates averaged over large numbers of molecular states. 

Ultracold experiments will much more accurately resolve the states involved.
This resolution is dramatically enhanced even further by an optical lattice: It can be several orders of magnitude more precise than the already spectacular resolution of an ultracold gas in the absence of a lattice. Sec.~\ref{sec:outlook} discusses this more.  

Developing theories that are able to describe the state-by-state variation can have an important impact on understanding and controlling chemical reactions, even those at room temperatures that ultimately average over many states. This is because upon varying a parameter (some molecular property or external field) away from a regime in which the TST describes the dynamics, the relevant average over states may be altered. Quoting Levine, ``One of the greatest challenges in chemistry is to $\ldots$ reveal how chemical transformations occur that are otherwise hidden behind thermal averages and multi-step mechanisms."~\cite{levine:molecular_2010}.

\subsection{Bimolecular complex polarizability \label{sec:complex-pol}}

In Refs.~\cite{wall:microscopic-derivation-multichannel_2016,docaj:ultracold_2016} it was assumed that the polarizability of the BCCs $\alpha_c$ was twice that of individual molecules $\alpha_m$.  The intuition for this estimate is that polarizability scales roughly linearly with the size of the object, and the complex states which contribute most at threshold energies are loosely bound, occupying a physical volume more or less twice that of a single molecule.  This is similar to loosely bound Feshbach molecules, which have a polarizability approximately twice that of their constituent atoms.  For example, the polarizability of Cs$_2$ in its least bound vibrational level has been predicted to be 1.96$\alpha_{\mathrm{Cs}}$ and measured to be $2.02\alpha_{\mathrm{Cs}}$~\cite{Vexiau2011,danzl2008quantum} and the polarizability of heteronuclear KRb Feshbach molecules has also been measured to be consistent with $\alpha_{\mathrm{K}}+\alpha_{\mathrm{Rb}}$~\cite{zirbel2008ultracold}.  For ground state molecules, which in the context of atomic scattering are somewhat analogous to of our closed-channel dominated BCC resonances in molecular scattering, the polarizability can be substantially different from twice the atomic polarizability, especially near resonances~\cite{1367-2630-17-6-065019}.

\begin{figure}
\includegraphics[width=0.75\columnwidth]{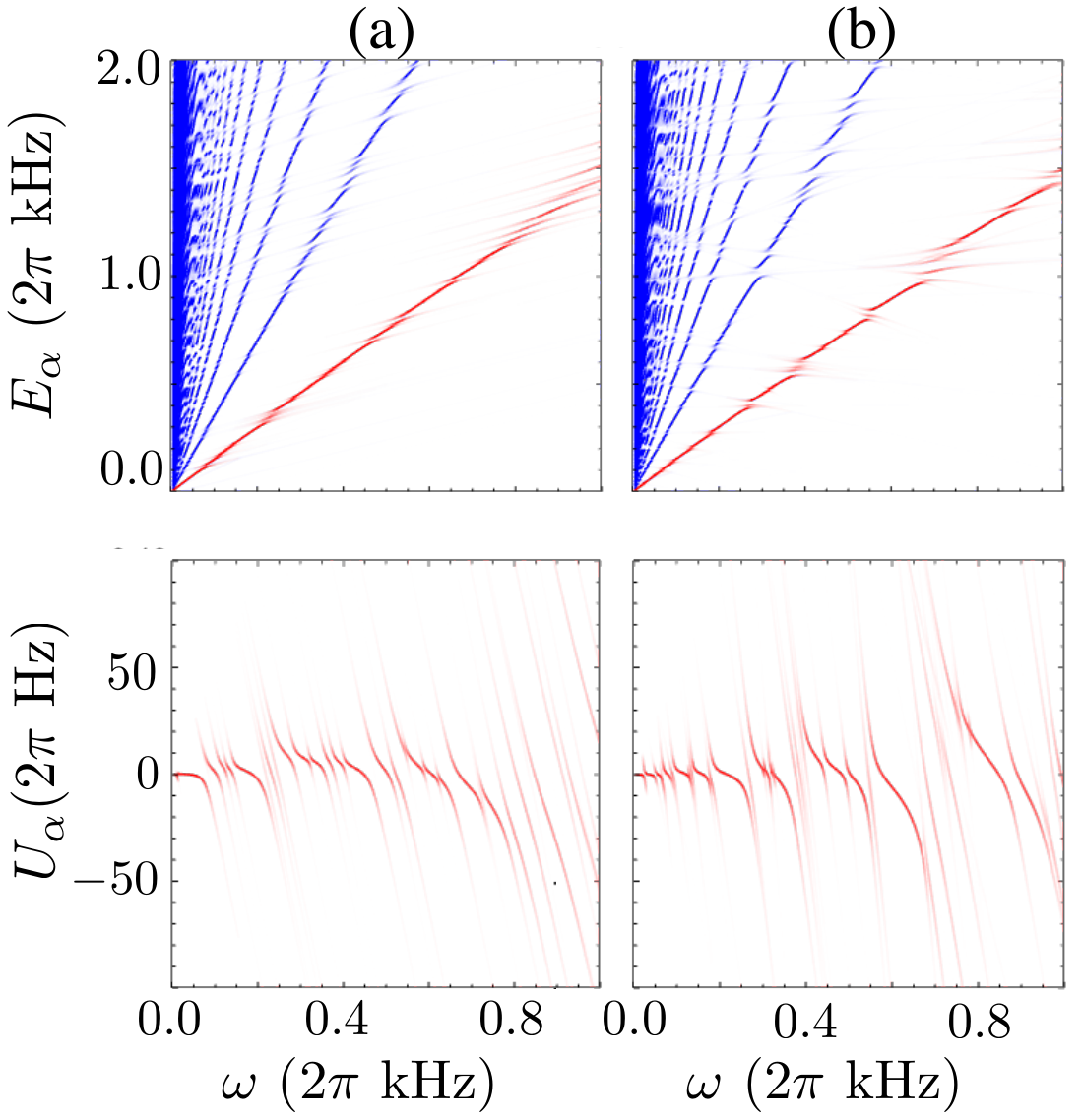}
\caption{\textbf{Effect of bimolecular collision complex ac polarizability deviating from twice the molecules' ac polarizability.} $E_\alpha$ (left) and $U_\alpha$ (right) versus trap depth $\omega$, with weight set as in Fig.~\ref{fig:beyond-RMT-model-props}. (a) Bimolecular collision complex polarizability $\alpha_c$ is taken to be equal to the single molecule polarizability $\alpha_m$ (rather than $\alpha_c \approx 2\alpha_m$ that was taken in earlier plots). (b) Each bimolecular collision complex has a polarizability independently sampled from a Gaussian with standard deviation $\alpha_m$.  
\label{fig:polarizability-disperse}}
\end{figure}

Let us now consider the physical consequences of $\alpha_c\ne 2\alpha_m$.  First, this leads to a different trapping frequency $\omega_c$ compared to the frequency $\omega$ of individual molecules, with $\omega_c/\omega=\sqrt{\alpha_c/2\alpha_m}$.  Here, the factor of two in the denominator comes from the fact that the BCCs have twice the mass of molecules.  Next, we note that the root-mean-square size of the closed channel wavefunction in the relative coordinate is $\langle R^2\rangle\sim \rsr^2$, and so the associated contribution of the finite size of the complex to the harmonic trapping energy is $\sim \omega_c (\rsr/\lho)^2$.  The ratio of the harmonic and short range length scales $\lho/\rsr\sim 25$ in typical situations, and so the relative coordinate contributes an energy $\sim 0.1\%$ of the zero-point energy, which we will neglect.  Hence, the only potential energy contribution arising from the BCCs comes from the center of mass motion, and takes the value
\begin{align}
\nonumber &\left(2n_{{\text{COM}}}+\ell_{{\text{COM}}}+3/2\right)\omega_c\\
&=\left(2n_{{\text{COM}}}+\ell_{{\text{COM}}}+3/2\right)\sqrt{\alpha_c/(2\alpha_m)}\omega\, .
\end{align}
In summary, the energies of the BCCs as a function of $\omega$ show a ``dispersion" with respect to free molecules, as shown in Fig.~\ref{fig:polarizability-disperse}.  Since our resonances are predicted only statistically, adding this dispersion causes no qualitative changes in our model at fixd $\omega$.  Instead, it just shifts the energies of the BCCs as a function of $\omega$.

Finally, we note that the difference in trapping frequencies introduces Franck-Condon factors $\mathcal{F}_{n_{{\text{BCC}}},\ell_{{\text{BCC}}};n_{{\text{COM}}},\ell_{{\text{COM}}}}=\langle n_{{\text{BCC}}},\ell_{{\text{BCC}}}|n_{{\text{COM}}},\ell_{{\text{COM}}}\rangle$ associated with a pair of molecules in the center of mass state $|n_{{\text{COM}}},\ell_{{\text{COM}}}\rangle$ making a transition to a BCC with center of mass state $| n_{{\text{BCC}}},\ell_{{\text{BCC}}}\rangle$.  As $\alpha_c\to 2\alpha_m$, $\mathcal{F}_{n_{{\text{BCC}}},\ell_{{\text{BCC}}};n_{{\text{COM}}},\ell_{{\text{COM}}}}\to \delta_{n_{{\text{BCC}}},n_{{\text{COM}}}}\delta_{\ell_{{\text{BCC}}},\ell_{{\text{COM}}}}$.  Generally speaking, these Franck-Condon factors will tend to further narrow closed-channel dominated resonances, for which $\alpha_c$ may be significantly different from $2\alpha_m$, and leave the width of open-channel dominated resonances relatively unaffected.

\section{Conclusions and outlook \label{sec:outlook}}

In summary, we have calculated the $U_\alpha$ and ${\mc O}_\alpha$ parameters that appear in the effective lattice model Eq.~\eqref{eq:EffectiveModel}, whose form was derived in Ref.~\cite{wall:microscopic-derivation-multichannel_2016}.  Because a full microscopic calculation of these parameters is intractable, we necessarily turned to approximations. 
As a first step, we calculated these within the same standard suite of approximations that were employed in Ref.~\cite{docaj:ultracold_2016}. 

Section~\ref{sec:approx} described this standard suite of approximations in considerable detail. These approximations consisted of RMT, QDT, TST, stitched together via the separation of length scales $\rsr \ll \rvdw \ll \lho$. In addition to a more extensive presentation of  Ref.~\cite{docaj:ultracold_2016}'s approximations, Sec.~\ref{sec:approx} furthermore derived the criteria for their applicability and discussed the likelihood that NRMs satisfy these criteria. 

While some of these approximations are widely applied, some are uncontrolled and one may expect some deviations, especially quantitative ones. Sec.~\ref{sec:approx} describes possible deviations from the standard suite of approximations, and presents more accurate theories to account for them. It also considers the consequences of these deviations for the $U_\alpha$s and ${\mc O}_\alpha$s. 

In light of these uncertainties, a key goal going forward will be to assess the accuracy of the predictions based on these various approximations.  Ultracold experiments have long been understood to provide an extremely tunable and high-accuracy system in which to probe the chemical behaviors represented in our approximations, largely by virtue of their low $\lsim \mu$K temperatures. In a lattice, this energy resolution is increased by several more orders of magnitude, to well under a nK. This is because the energy in a deep lattice is precisely quantized at the harmonic oscillator frequency; temperature, rather than smearing out the energy, transfers weight to other discrete frequencies that have little impact on the measurement of interest. As one example, experiments can utilize lattice modulation spectroscopy to probe the parameters $U_\alpha$ and ${\mc O}_\alpha$. Taking a rather pessimistic estimate of an interrogation time of $100$ms, well under the $\gsim10$s lifetimes already observed for \textit{reactive} molecules in a lattice, one finds a $0.5$nK energy resolution. Pushing this interrogation time towards potential lifetimes of nonreactive molecules in a  lattice would allow the lattice parameters and the approximations underlying them to be probed with energy resolution approaching a picoKelvin! 

Whether or not experiments measure properties in agreement with the  predictions for $U_\alpha$ and ${\mc O}_\alpha$, there will be a broad-reaching scientific payoff.  If the approximations are found to be accurate, then we will possess a quantitative lattice model suitable for future studies of NRMs in an optical lattice, analogous to the Hubbard model for ultracold atoms. This will open up qualitative new regimes of physics that are inaccessible with atoms. Furthermore, in this case approximations such as TST will be validated in an new regime with unprecedented accuracy. On the other hand, and arguably even more exciting, if experiments find the approximations are found to be lacking,  this will have impact beyond ultracold physics, highlighting the limits of these broadly used approximations and paving the way to forge new ones.

\section{Acknowledgements}

We thank Jose D'Incao, Kevin Ewart, Paul Julienne, and Brandon Ruzic
for useful discussions.  KRAH acknowledges the Aspen Center for Physics, which is supported by National Science Foundation grant PHY-1066293, for its hospitality while part of this work was performed. This work was supported with funds from the Welch foundation, Grant No. C-1872.  MLW acknowledges support from the NRC postdoctoral fellowship program.

\bibliography{NRM-1site-v7}

\end{document}